\documentclass[12pt,a4paper,english]{article}
\usepackage{titling}
\usepackage[affil-it]{authblk}
\usepackage{longtable}
\usepackage{lscape}
\usepackage{graphicx}
\usepackage{epsfig}
\usepackage{dsfont}
\usepackage{amsfonts}
\usepackage{amsmath}
\usepackage{amsthm}
\usepackage{amssymb}
\usepackage{bbold}
\usepackage[english]{babel}
\usepackage[utf8]{inputenc}
\usepackage[left=0.8in,right=0.8in,top=1.0in,bottom=1.2in]{geometry}
\usepackage{calc}
\usepackage{cancel}
\usepackage{float}
\usepackage{slashed}
\usepackage{mathrsfs}
\usepackage{pdfpages}
\usepackage{tabu}
\usepackage{simplewick}
\usepackage[hidelinks]{hyperref}
\usepackage{bm}
\usepackage{multirow}
\usepackage{hhline}
\usepackage{pifont}
\usepackage{float}
\usepackage[caption = false]{subfig}
\usepackage[normalem]{ulem}
\usepackage{cite}
\usepackage{tikz}
\usepackage{xcolor}

\usepackage{datetime}
\usepackage{booktabs}

\numberwithin{equation}{section}
\numberwithin{figure}{section}
\numberwithin{table}{section}

\makeatletter
\g@addto@macro\bfseries{\boldmath}
\makeatother

\allowdisplaybreaks

\definecolor{ForestGreen}{RGB}{34,139,34}

\newcommand{\Lb}{\Lambda_{b}}
\newcommand{\Lst}{\Lambda^*}
\newcommand{\mLb}{m_{\Lambda_{b}}}
\newcommand{\mLst}{m_{\Lambda^*}}
\newcommand{\sLb}{s_{\Lambda_b}}
\newcommand{\sL}{s_{\Lambda^*}}

\newcommand{\alphaLbLst}{\alpha_{\Lambda_b\Lambda^*}}
\newcommand{\schi}[3]{\left.\chi^{#1}_{#2}\right|_{#3}}

\makeatletter
\def\mb{\@ifstar\@@mb\@mb}
\newcommand{\@mb}[1]{\textcolor{red}{[\textbf{MB:} #1]}}
\newcommand{\@@mb}[1]{\textcolor{red}{#1}}

\def\ya{\@ifstar\@@ya\@ya}
\newcommand{\@ya}[1]{\textcolor{purple}{[\textbf{YA:} #1]}}
\newcommand{\@@ya}[1]{\textcolor{purple}{#1}}

\def\mr{\@ifstar\@@mr\@mr}
\newcommand{\@mr}[1]{\textcolor{green}{[\textbf{MR:} #1]}}
\newcommand{\@@mr}[1]{\textcolor{green}{#1}}
\makeatother

\def\eq#1{{Eq.~(\ref{#1})}}
\def\eqs#1#2{{Eqs.~(\ref{#1})--(\ref{#2})}}
\def\fig#1{{Fig.~\ref{#1}}}
\def\figs#1#2{{Figs.~\ref{#1}--\ref{#2}}}
\def\Table#1{{Table~\ref{#1}}}

\def\sec#1{{Sect.~\ref{#1}}}

\newcommand{\alignmentGrid}{
\draw[step=0.5cm,black,dotted,very thin](0,0)grid(\the\paperwidth,\the\paperheight);
\draw[step=1cm,black,dotted,thin](0,0)grid(\the\paperwidth,\the\paperheight);
\foreach \x in {0,2,4,6,8,10,12}
\draw (\x cm,1 pt)--(\x cm,-1 pt) node[anchor=south]{\color{black}$\x$};
\foreach \y in {2,4,6,8,10,12}
\draw (1 pt,\y)--(-1 pt,\y) node[anchor=west]{\color{black}$\y$};
}

\allowdisplaybreaks

\newcommand{\published}[1]{%
\gdef\puB{#1}}
\newcommand{\puB}{}

\def \L    {\Lambda}
\def \Lst  {\Lambda^*}
\def \Lb   {\Lambda_b}
\date{}
 
\begin{document}

\begin{titlepage}

\title{\textbf{Dispersive analysis of  $\Lb \to \L (1520)$ local form factors}}
\author[1]{Yasmine Amhis\thanks{yasmine.sara.amhis@cern.ch}}
\author[2]{Marzia Bordone\thanks{marzia.bordone@cern.ch}}
\author[3]{M\'eril Reboud\thanks{meril.reboud@tum.de}}
\affil[1]{Universit\'e Paris-Saclay, CNRS/IN2P3, IJCLab, Orsay, France}
\affil[2]{Theoretical Physics Department, CERN, 1211 Geneva 23, Switzerland}
\affil[3]{Technische Universit\"at M\"unchen, 
Physik Department, James-Franck-Stra\ss{}e 1, 85758 Garching}

\published{\flushright CERN-TH-2022-130, TUM-HEP-1412/22, EOS-2022-03\vskip2cm}

\maketitle
\thispagestyle{empty}

\begin{abstract}
We perform an analysis of $\Lb\to\L(1520)$ local form factors. We use dispersive techniques to provide a model-independent parametrisation of the form factors that can be used in the whole kinematic region. We use lattice QCD data to constrain the free parameters in the form factors expansion, which is further constrained by endpoint relations, dispersive bounds, and SCET relations. We analyse different scenarios, where we expand the form factors up to different orders, and their viability. Finally, we use our results to obtain predictions for some observables in $\Lb\to\L(1520)\ell^+\ell^-$ decays, as the differential branching ratio, the forward-backwards lepton asymmetry and the branching ratio of $\Lambda_b\to\L(1520)\gamma$. Finally, we provide a python notebook based on the software \texttt{EOS} to reproduce our result.
\end{abstract}

\end{titlepage}

\newpage

\clearpage

\section{Introduction}

Flavour Changing Neutral currents mediated through $b\to s\ell^+\ell^-$ transitions are an ideal laboratory to search for physics beyond the Standard Model (SM).
Their parametrically suppressed amplitudes, due to the GIM mechanism, render them a powerful probe of the intrinsic properties of the Standard Model itself, but also of the New Physics (NP) ones, should it be present.  

A key strategy in the study of this type of transitions consists in measuring a large number of observables in $b$-hadrons decays and seeing if patterns emerge. 
In the last decade, $B\to K^{(*)}\ell^+\ell^-$ decays received a lot of attention from both the theory and experimental communities. Indeed, the hadronic transitions $B\to K^{(*)}$ have been calculated using Lattice QCD techniques mostly at high $q^2$, where $q^2$ is the invariant mass of the dilepton system in the final state \cite{Horgan:2013hoa,Bouchard:2013eph,Horgan:2015vla,Bailey:2015dka,Parrott:2022dnu}, and also by means of other non-perturbative techniques such as Light-Cone Sum Rules \cite{Bharucha:2015bzk,Gubernari:2018wyi}. Experimentally, several measurements have been performed, from total branching fractions to  angular distributions \cite{LHCb:2020gog,LHCb:2020lmf,LHCb:2013zuf,ATLAS:2018gqc,CMS:2020oqb}. Furthermore, some tensions have been observed when comparing $B\to K^{(*)}\mu^+\mu^-$ and $B\to K^{(*)}e^+ e^-$ decays, at the level of $2.5-3.1$ standard deviations\cite{LHCb:2021lvy,LHCb:2019hip,LHCb:2017avl,LHCb:2014vgu}. In light of this, it is essential to assess $b$-baryon modes to expand our scope of understanding.\\
While the underlying quark-level transition remains the same with respect to mesons, decays of baryons have a different spin structure.
They, therefore, provide the possibility of studying a richer pattern of observables.
From a theoretical point of view, studies of local contributions to the ground state $\Lb\to\L$ transition have already been performed, both using Lattice QCD \cite{Detmold:2016pkz} and dispersive analysis \cite{Blake:2022vfl}, and measurements of $\Lambda_b\to\Lambda \mu^+\mu^-$ angular distribution are available in Ref.~\cite{LHCb:2018jna}.
However, from an experimental point of view, these decays are challenging to reconstruct because of the long lifetime {$\cal O$}(100 ps) of the weakly decaying $\L$ particle \cite{LHCb:2018jna}.
This is not the case of $\Lst$ excited states, such as the $\L(1520)$, which decays strongly. 
The difficulty in the treatment of $\Lst$ excited states arises from the presence of many overlapping resonant states that need to be disentangled experimentally. Despite this challenge, the $\L(1520)$ resonance has the advantage of being fairly narrow \cite{ParticleDataGroup:2020ssz} and produced in non-negligible quantities\cite{LHCb:2015yax}.

In this work, we perform an analysis of $\Lb\to\L(1520)$  local form factors.
This transition has already been studied using various techniques, which, however, are often applicable only in specific kinematic regions \cite{Descotes-Genon:2019dbw,Meinel:2021mdj,Meinel:2020owd,Bordone:2021bop,Mott:2011cx,Das:2020cpv}.
Up to now, no thorough study of an interpolation between the latter results that is valid in the whole kinematic region has been provided.
We address this precise issue, providing for the first time a parametrisation based on the analyticity of the form factors, viable in the entire phase space. In other words, our aim is to perform a pilot study to control the extrapolation uncertainties associated with local form factors.
To fix the coefficients of the form factors expansion, we use the results in Refs.~\cite{Meinel:2020owd,Meinel:2021mdj}, valid at low recoil (high $q^2$), endpoint relations at low and large recoil (low $q^2$), dispersive bounds and relations from Soft-Collinear Effective Theory.
We use our findings to perform numerical studies, providing the values for some observables that can be measured in the near future at the LHCb experiment \cite{Amhis:2020phx}.

This paper is organised as follows: in \sec{sec:parametrisation} we propose a form factors parametrisation based on the conformal mapping onto the complex $z$ plane, and we derive for the first time dispersive bounds for the $\Lb\to\L(1520)$ transition; in \sec{sec:fit} we discuss the fit of our parametrisation onto the available data; in \sec{sec:pheno}, we provide predictions for several observables; the conclusions of this work are drawn in \sec{sec:conclusions}.

\section{Theory framework}
\label{sec:parametrisation}
The $\Lb\to\L(1520)$ transition is characterised by $14$ hadronic form factors. We denote $\Lambda(1520)\equiv\Lst$ in the remainder of this work. The complete basis reads (using the conventions in Ref.~\cite{Meinel:2021mdj}): 
\begin{align}
 \nonumber \langle \Lst(k,\sL) | \bar{s} \,\gamma^\mu\, b | \Lambda_b(p,\sLb) \rangle =\,&+
 \bar{u}_\lambda(k,\sL)\bigg[ f_0(q^2)\, \frac{\mLst}{s_+} (\mLb-\mLst)\frac{q^\mu}{q^2} p^\lambda \\
 \nonumber & + f_+(q^2) \frac{\mLst}{s_-} \frac{\mLb+\mLst}{s_+}\left( p^\mu + k^{\mu} - (\mLb^2-\mLst^2)\frac{q^\mu}{q^2}  \right)p^\lambda \\
 &+ f_\perp(q^2) \frac{\mLst}{s_-} \left(\gamma^\mu - \frac{2\mLst}{s_+} p^\mu - \frac{2 \mLb}{s_+} k^{ \mu} \right)p^\lambda\nonumber\\
 &+ f_{\perp^\prime}(q^2) \frac{\mLst}{s_-} \left(\gamma^\mu p^\lambda + \frac{2(\mLb k^\mu+\mLst p^\mu)}{s_+} p^\lambda - \frac{2 k^\mu}{\mLst} p^{\lambda} \right. \nonumber\\
 &+ \left. \frac{s_-}{\mLst}g^{\lambda\mu} \right) \bigg] u_{\Lambda_b}(p,\sLb), \label{eq:HMEL1}\\
 \nonumber \langle \Lst(k,\sL) | \bar{s} \,\gamma^\mu\gamma_5\, b | \Lambda_b(p,\sLb) \rangle =\,&
 -\bar{u}_\lambda(k,\sL) \:\gamma_5 \bigg[ g_0(q^2)\:\frac{\mLst}{s_-} (\mLb+\mLst)\frac{q^\mu}{q^2}p^\lambda \\
 \nonumber & + g_+(q^2)\frac{\mLst}{s_+}\frac{\mLb-\mLst}{s_-}\left( p^\mu + k^{\mu} - (\mLb^2-\mLst^2)\frac{q^\mu}{q^2}  \right)p^\lambda \nonumber\\
 &+ g_\perp(q^2) \frac{\mLst}{s_+}\left(\gamma^\mu + \frac{2\mLst}{s_-} p^\mu - \frac{2 \mLb}{s_-} k^{\mu} \right)p^\lambda\nonumber\\
 &+ g_{\perp^\prime}(q^2)\frac{\mLst}{s_+} \left(\gamma^\mu p^\lambda + \frac{2(\mLb k^\mu-\mLst p^\mu)}{s_-} p^\lambda + \frac{2 k^\mu}{\mLst} p^{\lambda} \right. \nonumber \\
 &- \left. \frac{s_+}{\mLst}g^{\lambda\mu} \right) \bigg]  u_{\Lambda_b}(p,\sLb)\,, \label{eq:HMEL2}\\
\langle\Lst(k,\sL)|\bar{s} i \sigma^{\mu\nu} q_\nu b|\Lambda_b(p,s_{\Lambda_b})\rangle= \, & - \frac{\mLst}{s_-}\bar{u}_{\lambda}(k, s_\Lambda) \Big\{h_+(q^2)\frac{q^2 (p^\mu+k^{ \mu})-(\mLb^2-\mLst^2)q^\mu}{s_{+}}p^\lambda \nonumber\\
&+h_\perp (q^2)(\mLb+\mLst)\Big[\gamma^\mu-2\frac{\mLb k^\mu+\mLst p^\mu}{s_+} \Big]p^\lambda \nonumber\\
&+ h_{\perp^\prime}(\mLb+\mLst)\Big[\gamma^\mu p^\lambda+2\frac{\mLb k^\mu+\mLst p^\mu}{s_+}p^\lambda-2\frac{k^\mu}{\mLst}p^\lambda \nonumber\\
&+\frac{s_-}{\mLst}g^{\mu\lambda} \Big]\Big\} u_{\Lambda_b}(p, s_{\Lambda_b})\, \label{eq:HMEL3} \\
\langle\Lst(k,\sL)|\bar{s} i \sigma^{\mu\nu}q_\nu \gamma_5 b|\Lambda_b(p,s_{\Lambda_b})\rangle= \, & -\frac{\mLst}{s_+} \bar{u}_{\lambda}(k, s_\Lambda)\gamma_5 \Big\{\tilde{h}_+ \frac{q^2 (p^\mu+k^{ \nu})-(\mLb^2-\mLst^2)q^\nu}{s_{-}}p^\lambda\nonumber\\
&+\tilde{h}_\perp(q^2)(\mLb-\mLst) \Big[\gamma^\mu-2\frac{\mLb k^\mu-\mLst p^\mu}{s_-}\Big]p^\lambda\nonumber\\
&+\tilde{h}_{\perp^\prime}(\mLb-\mLst) \Big[\gamma^\mu p^\lambda+2\frac{\mLb k^\mu-\mLst p^\mu}{s_-}p^\lambda+2 \frac{k^\mu}{\mLst}p^\lambda\nonumber\\
&-\frac{s_+}{\mLst}g^{\lambda\mu}\Big]\Big\} u_{\Lambda_b}(p, s_{\Lambda_b})\,, \label{eq:HMEL3}
\end{align}
where $u_\lambda$ is the $3/2^-$ projection of a Rarita-Schwinger object \cite{Falk:1991nq}, and $\{p,\sLb\}$ and $\{k,\sL\}$ are the momenta and spin of the $\Lambda_b$ and $\Lst$, respectively.
For convenience, we also define the momentum transfer $q = p-k$, and $s_\pm = (\mLb\pm\mLst)-q^2$.\\
The matrix elements that we listed above obey endpoint relations at $q^2 = q^2_\mathrm{max} = (\mLb - \mLst)^2$ and $q^2=0$.
In our basis, these relations read\cite{Hiller:2021zth,Descotes-Genon:2019dbw,Meinel:2021mdj,Papucci:2021pmj}:
\begin{align}
    f_\perp(q^2_\mathrm{max}) =&~ - f_{\perp^\prime}(q^2_\mathrm{max})\,, \label{eq:fperpq2max} \\
    f_+(q^2_\mathrm{max})     =&~ 2 \, \frac{\mLb-\mLst}{\mLb+\mLst}f_{\perp^\prime}(q^2_\mathrm{max})\,, \\
    g_0(q^2_\mathrm{max})     =&~ 0 \,, \\
    g_+(q^2_\mathrm{max})     =&~ g_\perp(q^2_\mathrm{max})-g_{\perp^\prime}(q^2_\mathrm{max})\,, \label{eq:gpatq2max} \\
    h_\perp(q^2_\mathrm{max}) =&~ -h_{\perp^\prime}(q^2_\mathrm{max})\,, \\
    h_+(q^2_\mathrm{max})     =&~ 2 \, \frac{\mLb+\mLst}{\mLb-\mLst}h_{\perp^\prime}(q^2_\mathrm{max})\,, \\
    \tilde{h}_+(q^2_\mathrm{max}) =&~ \tilde h_{\perp}(q^2_\mathrm{max})- \tilde h_{\perp^\prime}(q^2_\mathrm{max})\,,\label{eq:htPlusq2max}
\end{align}
and
 \begin{align}
    f_0(0) =& \left(\frac{\mLb+\mLst}{\mLb-\mLst}\right)^2f_+(0)\,,  \label{eq:f0at0}\\
    g_0(0) =& \left(\frac{\mLb-\mLst}{\mLb+\mLst}\right)^2g_+(0)\,, \label{eq:g0at0}\\
    \tilde{h}_\perp(0)=& \left(\frac{\mLb+\mLst}{\mLb-\mLst}\right)^2 h_\perp(0)\,, \\
    -\tilde{h}_{\perp^\prime}(0)=& \left(\frac{\mLb+\mLst}{\mLb-\mLst}\right)^2 h_{\perp^\prime}(0)\,. \label{eq:htppat0}
\end{align}
In the following, we discuss the parametrisation that we adopt for the form factors and how they are subject to bounds from dispersive analysis.

\subsection{Form factors parametrisation}
\begin{figure}
    \centering
    \includegraphics{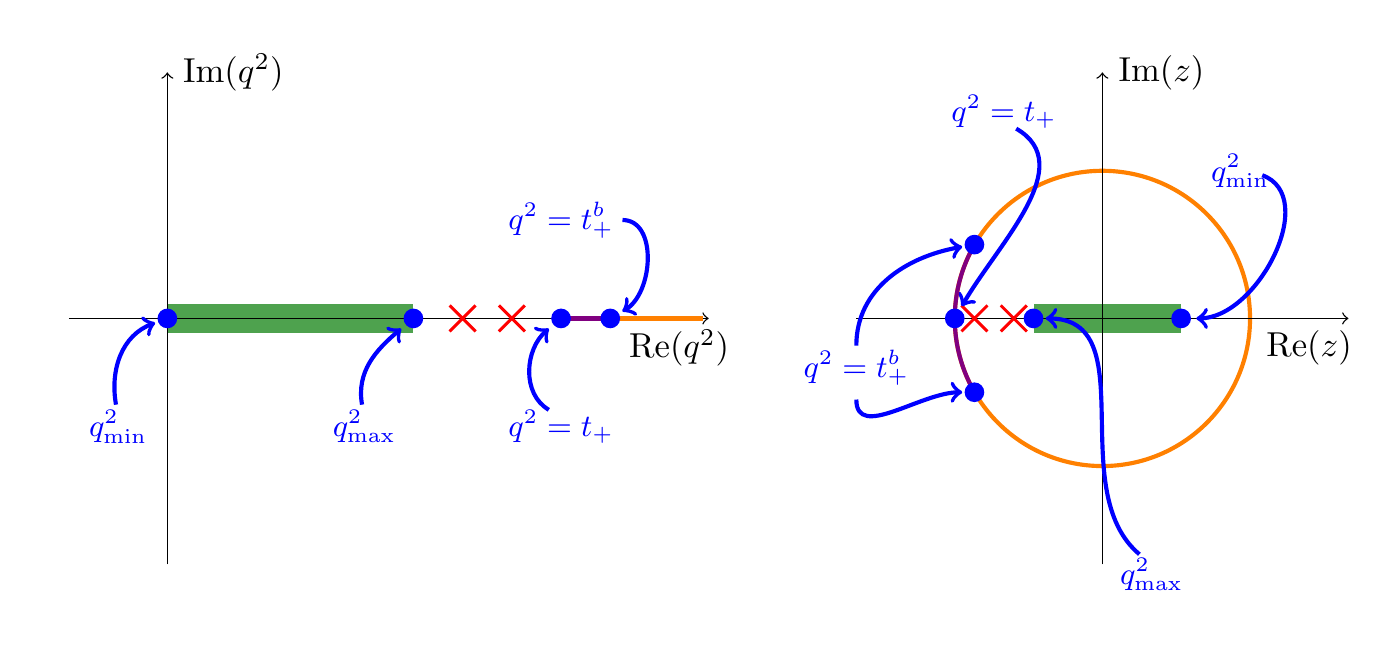}
    \caption{Illustration of the complex $q^2$ and $z$ plane. The green region represents the semileptonic region, the red crosses possible sub-threshold resonances, and the orange contours the branch cut. The purple line(arc) in the left(right) panel arises from $t_+\neq t^b_+$. In blue we indicate the relevant $q^2$ points.}
    \label{fig:z_plane}
\end{figure}

The analytic properties of the form factors suggest that it is convenient to parametrise them in terms of the complex variable $z$, that is defined by the conformal transformation
\begin{equation}
\label{eq:z_ef}
z(q^2, t_0) = \frac{\sqrt{t_+-q^2}-\sqrt{t_+-t_0}}{\sqrt{t_+-q^2}+\sqrt{t_+-t_0}}\,,
\end{equation}
which maps the $q^2$ axis onto the complex $z$ plane, with the advantage that $|z|\leq1$.
We set the parameter $t_+$ to be the two-body threshold production for $\bar{s}b$ states, which is $t_+ = (m_B+m_K)^2$.
We also choose $t_0 = t_+ - \sqrt{t_+^2 - t_+ t_-} = 9.865\, \text{GeV}^2$, with $t_-=(m_B-m_K)^2$, to lower the maximum allowed values for $z$ in the semileptonic region to $|z_\text{max}|\sim 0.09$. For convenience, we further introduce the quantities $t_{\pm}^b = (\mLb \pm \mLst)^2$. \\
This setup has been used to work out the so-called BGL parametrisation \cite{Boyd:1994tt,Boyd:1995cf}, whose success lies in the model independency of the parametrisation itself.
However, for transitions in which $t_+$ does not coincide with the threshold productions of the hadrons involved, this formalism needs to be slightly revisited in order to account for the fact that the region where $q^2>t_+$ is mapped only on an arc of the unit circle described by $|z|=1$, as it will be clarified at a later stage. This formalism has been introduced in Refs.~\cite{Blake:2022vfl,Gubernari:2020eft}, and we closely follow that approach.
The analytic properties of the form factors boil down to a branch-cut starting at $q^2=t_+$, which is the first hadronic threshold contribution, and potential sub-threshold contributions. The former is mapped on an arc of the unit circle, while the latter are on the real $z$ axis, as illustrated in \fig{fig:z_plane}.
Once we factor out these sub-threshold resonances in the form factors by introducing the so-called Blaschke factor, the remaining function is analytic inside the unit disk.
Hence, we parametrise the form factors as follows:
\begin{equation}
f_i(z) = \frac{1}{P_{f_i}(z)\phi_{f_i}(z)}\sum_{n=0}^N a^{f_i}_n \, p_n(z)\,,
\label{eq:z_exp}
\end{equation}
where $N$ is the truncation order of the series, and $p_n$ are orthonormal polynomials on the arc of the unit circle, as described in Appendix \ref{app:orthonormal_polynomials}. The Blaschke factor $P_{f_i}(z)$ is written as
\begin{equation}
    P_{f_i}(z)= \prod_j \frac{z - z(m_j^2, t_0)}{1 - z \, z^*(m_j^2, t_0)}
\end{equation}
where $m_j$ are the masses of the sub-threshold resonances listed in \Table{tab:resonances} and $z^*$ is the complex conjugate of $z$. Finally, the functions $\phi_{f_i}$ are the so-called outer functions, for which we give an explicit representation in the next Section.
The advantage of this setup is that it allows to apply dispersive techniques to bound the form factors coefficients in \eq{eq:z_exp}, as illustrated in the next Section.

\begin{table}
\begin{center}
\renewcommand{\arraystretch}{1.2} 
\begin{tabular}{ c c c c c}
\toprule
Form factor & $J^P$ & Resonance & $m_\text{res}^j [\mathrm{GeV}]$ & $\schi{J}{\Gamma}{\mathrm{OPE}}\times 10^2$ \\
\midrule
$g_0$& $0^-$ & $B_s$ & $5.367$\, \cite{ParticleDataGroup:2020ssz} & $1.57$\\
$g_{+}$, $g_{\perp}$, $g_{\perp^\prime}$ & $1^+$ & $B_{s,1}$ & $5.750$\, \cite{Lang:2015hza} & $1.13/m_b^2$\\
$f_0$& $0^+$ & $B_{s,0}$ & $5.711$\, \cite{Lang:2015hza} & $1.42$\\
$f_+$, $f_{\perp}$, $f_{\perp^\prime}$& $1^-$ & $B_s^*$ & $5.416$\, \cite{ParticleDataGroup:2020ssz} & $1.20/m_b^2$\\
$h_{+}$, $h_{\perp}$, $h_{\perp^\prime}$& $1^-$ & $B_s^*$ & $5.416$\, \cite{ParticleDataGroup:2020ssz} & $0.803/m_b^2$\\
$\tilde{h}_{+}$\, $\tilde{h}_{\perp}$, $\tilde{h}_{\perp^\prime}$& $1^+$ & $B_{s,1}$ & $5.750$\, \cite{Lang:2015hza} & $0.748/m_b^2$\\
\toprule
\end{tabular}
\caption{Parameters for the form factor expansion. The reference value for the $b$-quark mass is $m_b=4.2\,\text{GeV}$.}
\label{tab:resonances}
\end{center}
\end{table}

\subsection{Dispersive Bounds}

The main idea is to use the dispersive analysis to extract bounds on the form factors parameters \cite{Caprini:2019osi,Boyd:1994tt,Boyd:1995cf}. In this, we follow closely the approach of Ref.~\cite{Blake:2022vfl}. The starting point is the two-point correlation function:
\begin{align}
    \Pi^{\mu\nu}_{\Gamma}(q)
        & \equiv i \int d^4x\, e^{iq\cdot x} \langle 0 | \mathcal{T}[J_\Gamma^\mu(x), J_\Gamma^{\dagger,\nu}(0)] | 0\rangle
         = \left(\frac{q^\mu q^\nu}{q^2} - g^{\mu\nu}\right) \Pi_{1}^\Gamma(q^2) + \frac{q^\mu q^\nu}{q^2} \Pi_{0}^\Gamma(q^2)
        \,,
\end{align}
where $\Gamma = V,A,T,T5$ for vector, axial-vector, tensor and pseudo-tensor currents, respectively. We note that  $\Pi_{0}^\Gamma(q^2)=0$ for tensor and pseudo-tensor currents. We introduce the projectors:
\begin{equation}
    \mathcal{P}^0_{\mu\nu} = \frac{q_\mu q_\nu}{q^2}\,, \qquad \mathrm{and} \qquad \mathcal{P}^1_{\mu\nu} = \frac{1}{3}\left(\frac{q_\mu q_\nu}{q^2}-g_{\mu\nu}\right)\,,
\end{equation}
such that
\begin{equation}
    \mathcal{P}^0_{\mu\nu}\,\Pi^{\mu\nu}_{\Gamma}(q) = \Pi_{0}^\Gamma(q^2)\,, \qquad \mathrm{and} \qquad \mathcal{P}^1_{\mu\nu}\,\Pi^{\mu\nu}_{\Gamma}(q) = \Pi_{1}^\Gamma(q^2)\,.
\end{equation}
To evaluate the bounds, we are actually interested in extracting the discontinuity of $\Pi^{\mu\nu}_\Gamma$. Therefore, we define the susceptibility
\begin{equation}
    \chi_\Gamma^J (Q^2) = \frac{1}{n!}\left(\frac{d}{dQ^2}\right)^n \Pi_\Gamma^J(Q^2) =\frac{1}{2\pi i}\int_0^\infty dt \,\frac{\mathrm{ Disc}[\Pi_V^J(t)]}{(t-Q^2)^{n+1}}\,,
    \label{eq:chi}
\end{equation}
where $n$ is the number of subtractions and might vary for the different currents. The bounds are obtained using quark-hadron duality, which implies
\begin{equation}
    \schi{J}{\Gamma}{\mathrm{OPE}} =  \schi{J}{\Gamma}{\mathrm{hadr}}\,,
    \label{eq:QHD}
\end{equation}
where the left-hand side (l.h.s.) is the perturbative calculation obtained in the Operator-Product-Expansion (OPE). The results for each current and angular momentum have been calculated in Ref.~\cite{Bharucha:2010im} and include $\mathcal{O}(\alpha_s)$ and $\mathcal{O}(\Lambda_\mathrm{QCD}/m_b)$ corrections, and are reported in \Table{tab:resonances}. In the right-hand side (r.h.s.) of \eq{eq:QHD}, one uses the spectral representation to introduce a sum over possible states mediated by an hadronic $\bar{s}b$ state. This yields:
\begin{equation}
    \mathrm{Disc}[\Pi_V^J(t)]= i \sum_{\mathrm{spin}}\int \mathrm{d}\rho (2\pi^4)\delta^{(4)}(q-\sum_{i=1}^n p_i)\mathcal{P}_{\mu\nu}^J\langle 0|J^\mu_\Gamma|H_{\bar{s}b}(p_1,\dots p_n)\rangle\langle H_{b\bar{s}}(p_1,\dots p_n)|J_\Gamma^{\nu\dagger}|0\rangle\,,
\end{equation}
that is plugged in \eq{eq:chi} to obtain $\schi{J}{\Gamma}{\mathrm{hadr}}$. We can use the number of states in $H_{\bar{s}b}(p_1,\dots p_n)$ to further decompose $\schi{J}{\Gamma}{\mathrm{hadr}}$, finding
\begin{equation}
    \schi{J}{\Gamma}{\mathrm{hadr}} = \schi{J}{\Gamma}{\mathrm{1pt}} +  \schi{J}{\Gamma}{\mathrm{2pt}} + \dots\,,
    \label{eq:chihad_decomp}
\end{equation}
where the first term in the r.h.s. of \eq{eq:chihad_decomp} encodes the one particle contribution, the second term the two-particle contribution, and the ellipses stand for higher multiplicity states. The one-particle contribution accounts for the $B_s$ and $B_s^*$ states. For each current and angular momentum state, we find
\begin{align}
    \schi{1}{V}{\mathrm{1pt}} =& ~ \frac{m_{B_s^*}^2 f_{B_{s}^*}^2}{(m_{B_s^*}^2-Q^2)^{n+1}}\,, & \schi{0}{V}{\mathrm{1pt}} =& ~ \frac{m_{B_{s,0}^*}^2 f_{B_{s,0}^*}^2}{(m_{B_{s,0}^*}^2-Q^2)^{n+1}}\,, \\
    \schi{1}{A}{\mathrm{1pt}} =& ~ \frac{m_{B_{s,1}^*}^2 f_{B_{s,1}^*}^2}{(m_{B_{s,1}^*}^2-Q^2)^{n+1}}\,, & \schi{0}{A}{\mathrm{1pt}} =& ~ \frac{m_{B_{s}}^2 f_{B_{s}}^2}{(m_{B_{s}}^2-Q^2)^{n+1}}\,, \\
    \schi{1}{T}{\mathrm{1pt}} =& ~ \frac{m_{B_{s}^*}^4 (f_{B_{s,1}^*}^T)^2}{(m_{B_{s}^*}^2-Q^2)^{n+1}}\,, & \schi{1}{T5}{\mathrm{1pt}} =& ~ \frac{m_{B_{s,1}}^4 (f_{B_{s,1}}^T)^2}{(m_{B_{s}}^2-Q^2)^{n+1}}\,,
\end{align}
in agreement with Ref.~\cite{Blake:2022vfl}. \\
The two-particle contribution in principle contains an infinite sum over two-body hadronic states (e.g. $\bar{B}K$, $\bar{B}K^*$, $\bar{B}_s\phi$, etc.).
It is far beyond the scope of this work to perform a global analysis of $b\to s$ transition; therefore, in the following, we only consider $H_{\bar{s}b}(p_1,\dots p_n)=\Lambda_b(p_1)\bar\Lambda^*(-p_2)$.
With this choice, we find 
\begin{align}
\label{eq:dispersive_bounds_V0}
    \schi{0}{V}{\mathrm{2pt}} =&\, \frac{1}{48\pi^2}\int_{t_+^b}^\infty dt \frac{\sqrt{s_+ s_-^3}}{t^2(t-Q^2)^{n+1}}(\mLb-\mLst)^2|f_0|^2  \,, \\
    \schi{1}{V}{\mathrm{2pt}} =&\, \frac{1}{144\pi^2} \int_{t_+^b}^\infty dt \frac{\sqrt{s_+^3 s_-}}{t^2(t-Q^2)^{n+1}}\left[(\mLb+\mLst)^2 |f_+|^2+2t|f_\perp|^2+6t|f_{\perp^\prime}|^2\right] \,, \\
    \schi{0}{A}{\mathrm{2pt}} =&\,\frac{1}{48\pi^2} \int_{t_+^b}^\infty dt \frac{\sqrt{s_- s_+^3}}{t^2(t-Q^2)^{n+1}}(\mLb+\mLst)^2|g_0|^2  \,, \\
    \schi{1}{A}{\mathrm{2pt}} =&\, \frac{1}{144\pi^2} \int_{t_+^b}^\infty dt \frac{\sqrt{s_-^3 s_+}}{t^2(t-Q^2)^{n+1}}\left[(\mLb-\mLst)^2 |g_+|^2+2t|g_\perp|^2+6t|g_{\perp^\prime}|^2\right]  \,, \\
    \schi{1}{T}{\mathrm{2pt}} =&\,  \frac{1}{144\pi^2} \int_{t_+^b}^\infty dt \frac{\sqrt{s_+^3 s_-}}{t (t-Q^2)^{n+1}}\left[t |h_+|^2+2(\mLb+\mLst)^2|h_\perp|^2+6(\mLb+\mLst)^2|h_{\perp^\prime}|^2\right] \,, \\
\label{eq:dispersive_bounds_T51}
    \schi{1}{T5}{\mathrm{2pt}} =&\, \frac{1}{144\pi^2} \int_{t_+^b}^\infty dt \frac{\sqrt{s_-^3 s_+}}{t (t-Q^2)^{n+1}}\left[t |\tilde{h}_+|^2+2(\mLb-\mLst)^2|\tilde{h}_\perp|^2+6(\mLb-\mLst)^2|\tilde{h}_{\perp^\prime}|^2\right]  \,.
\end{align}
where $s_\pm = t_\pm^b - t$. The parameter $Q^2$ is chosen to ensure that the two-point function is highly virtual \cite{Boyd:1997kz}. In this context this means $Q^2-(m_b+m_s)^2\ll-(m_b+m_s)\Lambda_\mathrm{QCD}$, that is well respected if $Q^2=0$. Hence, we set $Q^2=0$ in the remainder of this work. \\
In order to formulate the bounds, we still need to parametrise the outer functions. Using the results in Refs.~\cite{Blake:2022vfl,Caprini:2019osi}, we have
\begin{equation}
\begin{aligned}
    \phi_{f_i} (z)&= \frac{(t_+^b)^{A/2}(t_-^b)^{B/2}(1-z)^{n+g-e/2-f/2-3/2}(1+z)^{1/2}}{\sqrt{16 d \,\pi^2 \schi{J}{\Gamma}{\mathrm{OPE}}}}\\
    &\times\left[\frac{-1}{t_0 (1 + z)^2 - 2 t_+ (1 + z^2) - 2 (1 - z^2) \sqrt{t_+} \sqrt{t_+ - t_0}}\right]^{(n+g)/2} \\
    &\times\left[- (1-z)^2 t_-^b-(1+z)^2 t_0+2t_+(1+z^2)+2(1-z^2)\sqrt{t_+ - t_-^b} \sqrt{t_+ - t_0}\right]^{e/4} \\ &\times\left[(1-z)^2 t_+^b-(1+z)^2 t_0+4 t_+z\right]^{f/4}\sqrt{4(t_+-t_0)}\,,
\end{aligned}
\end{equation}
where the coefficients $\{A,B,d,e,f,g,n\}$ are given in \Table{tab:coeff_phi}. Our representation of the outer functions is not unique, but their modulus square is fixed by \eqs{eq:dispersive_bounds_V0}{eq:dispersive_bounds_T51}. Any modification that preserves the analytic properties of $\phi_{f_i}$ and conserves $|\phi_{f_i}|^2$ is valid.\\
The relations in \eq{eq:chihad_decomp} are never exactly satisfied since we are unable to calculate all perturbative orders in the OPE, and we cannot sum over all possible hadronic states in the spectral representation. Taking this limitation into account, we obtain the inequality 
\begin{equation}
    \frac{\schi{J}{\Gamma}{\mathrm{hadr}}}{\schi{J}{\Gamma}{\mathrm{OPE}}} < 1\,,
\end{equation}
that, after imposing our previous findings, yields:
\begin{equation}
\label{eq:bounds_1}
\begin{aligned}
    &\sum_{i=0}^N |a^{f_0}_i|^2 <1-\frac{\schi{0}{V}{\mathrm{1pt}}}{\schi{0}{V}{\mathrm{OPE}}}\,, \qquad \sum_{i=0}^N (|a_i^{f_+}|^2+|a_i^{f_\perp}|^2+|a_i^{f_{\perp^\prime}}|^2) <1-\frac{\schi{1}{V}{\mathrm{1pt}}}{\schi{1}{V}{\mathrm{OPE}}}\,,\\
    &\sum_{i=0}^N |a^{g_0}_i|^2 <1-\frac{\schi{0}{A}{\mathrm{1pt}}}{\schi{0}{A}{\mathrm{OPE}}}\,, \qquad \sum_{i=0}^N (|a_i^{g_+}|^2+|a_i^{g_\perp}|^2+|a_i^{g_{\perp^\prime}}|^2) <1-\frac{\schi{1}{A}{\mathrm{1pt}}}{\schi{1}{A}{\mathrm{OPE}}}\,, \\
    &\sum_{i=0}^N (|a_i^{h_+}|^2+|a_i^{h_\perp}|^2+|a_i^{h_{\perp^\prime}}|^2) <1-\frac{\schi{1}{T}{\mathrm{1pt}}}{\schi{1}{T}{\mathrm{OPE}}}\,, \\
    &\sum_{i=0}^N (|a_i^{\tilde{h}_+}|^2+|a_i^{\tilde{h}_\perp}|^2+|a_i^{\tilde{h}_{\perp^\prime}}|^2) <1-\frac{\schi{1}{T5}{\mathrm{1pt}}}{\schi{1}{T5}{\mathrm{OPE}}}\,. 
\end{aligned}
\end{equation}
The above relations are usually addressed as dispersive bounds, and their simple form is a result of using the form factor parametrisation in \eq{eq:z_exp}.
The l.h.s. of these inequalities corresponds to the relative saturation of the bound due to two-particle contributions. In principle, one could add to \eq{eq:bounds_1} contributions from further two-body states and, in general, higher multiplicity states. This would render the bounds on the form factor coefficients that we are interested in extracting even stronger.

\begin{table}
\begin{center}
\renewcommand{\arraystretch}{1.1} 
\begin{tabular}{ c c c c c c c c }
\toprule
Form factor & $A$ &  $B$ & $d$ &  $e$ &  $f$ & $g$ & $n$ \\
\midrule
$f_0$ & $0$ & $1$ & $6$ & $3$ & $1$ & $3$ & $1$  \\
$f_+$ & $1$ & $0$ & $18$ & $1$ & $3$ & $3$ & $2$ \\
$f_\perp$ & $0$ & $0$ & $9$ & $1$ & $3$ & $2$ & $2$ \\
$f_{\perp^\prime}$ & $0$ & $0$ & $3$ & $1$ & $3$ & $2$ & $2$ \\
$g_0$ & $1$ & $0$ & $6$ & $1$ & $3$ & $3$ & $1$ \\
$g_+$ & $0$ & $1$ & $18$ & $3$ & $1$ & $3$ & $2$ \\
$g_\perp$ & $0$ & $0$ & $9$ & $3$ & $1$ & $2$ & $2$ \\
$g_{\perp^\prime}$ & $0$ & $0$ & $3$ & $3$ & $1$ & $2$ & $2$ \\
$h_+$ & $0$ & $0$ & $18$ & $1$ & $3$ & $1$ & $3$ \\
$h_\perp$ & $1$ & $0$ & $9$ & $1$ & $3$ & $2$ & $3$ \\
$h_{\perp^\prime}$ & $1$ & $0$ & $3$ & $1$ & $3$ & $2$ & $3$ \\
$\tilde{h}_+$ & $0$ & $0$ & $18$ & $3$ & $1$ & $1$ & $3$  \\
$\tilde{h}_\perp$ & $0$ & $1$ & $9$ & $3$ & $1$ & $2$ & $3$ \\
$\tilde{h}_{\perp^\prime}$ & $0$ & $1$ & $3$ & $3$ & $1$ & $2$ & $3$ \\
\toprule
\end{tabular}
\caption{Coefficients involved in the outer functions.}
\label{tab:coeff_phi}
\end{center}
\end{table}

\section{Fit Results}
\label{sec:fit}
We extract the coefficients in \eq{eq:z_exp} by fitting them against a set of data. They include
\begin{description}
    \item[Lattice QCD:] we use the results of Ref.~\cite{Meinel:2021mdj} to extract pseudo-points close to the position of the lattice points, namely at $q^2= 16.2975\,\mathrm{GeV}^2$ and $q^2=16.5537\,\mathrm{GeV}^2$. Two points per form factor are available at very low-recoil, yielding 28 pseudo-points. However, the results in Ref.~\cite{Meinel:2021mdj} take already into account endpoint relations at $q^2_\mathrm{max}$, effectively reducing the number of independent points to $21$.
    \item[Dispersive Bounds:] we apply the dispersive bounds in \eq{eq:bounds_1}. To account for the uncertainty in the computation of $\schi{J}{\Gamma}{\mathrm{OPE}}$, we implement the same approach of Ref.~\cite{Bordone:2019vic}, with a penalty function
    \begin{equation}
        -2 \log P(r)=
        \begin{cases}
            0 & \mathrm{if}\, r<1\,, \\
            \frac{(r-1)^2}{\sigma^2} & \mathrm{otherwise}\,,
        \end{cases}
    \end{equation}
    where $r$ is the saturation of the dispersive bounds, and we use  $\sigma=10\%$. This matches the precision of the OPE calculation in Ref.~\cite{Bharucha:2010im}.
    \item[Endpoint relations:] we apply endpoint relations at $q^2=0$ as in \eqs{eq:f0at0}{eq:htppat0} and at $q^2_\mathrm{max}$ as in \eqs{eq:fperpq2max}{eq:htPlusq2max}. Practically, these relations lower the number of free parameters, and seven of them are already accounted for in the lattice QCD extrapolation. We stress that endpoint relations are derived from fundamental properties of the hadronic amplitudes and are therefore independent of any chosen form factors model.
    \item[SCET Relations:] we use the fact that at large recoil Soft-Collinear Effective Theory (SCET) predicts the form factors to be either zero or proportional to a single soft form factor at leading order in $\alpha_s/\pi$ and leading power in $\Lambda_\mathrm{QCD}/m_b$ \cite{Boer:2014kda,Feldmann:2011xf,Mannel:2011xg,Descotes-Genon:2019dbw}.
    These relations are needed due to the lack of computation either on the lattice or in Sum-Rule approaches at large recoil.
    We apply these relations at $q^2=0$ with an uncertainty of $0.2$ that accounts for the unknown next-to-leading order and next-to-leading power corrections.
    In principle, this uncertainty is strongly correlated among all form factors.
    However, the correlations are impossible to guess without explicit computation of higher-order effects in the SCET expansion.
    To circumvent this problem, we restrict ourselves to the following set of relations:
    \begin{equation}
    \begin{aligned}
        f_{\perp'}(0) &=\, 0 \pm 0.2 \,, &  g_{\perp'}(0) &=\, 0 \pm 0.2 \,, &
        h_{\perp'}(0) &=\, 0 \pm 0.2 \,, \\
        \tilde{h}_{\perp'}(0) &=\, 0 \pm 0.2  \,, & f_+(0) / f_\perp(0) &=\, 1 \pm 0.2 \,, &
        f_\perp(0) / g_0(0) &=\, 1 \pm 0.2 \,, \\
        g_\perp(0) / g_+(0) &=\, 1 \pm 0.2 \,, &  h_+(0) / h_\perp(0) &=\, 1 \pm 0.2 \,, & f_+(0) / h_+(0) &=\, 1 \pm 0.2\,,
    \end{aligned}
    \end{equation} 
    and we treat the uncertainties as uncorrelated. This corresponds to a conservative assumption, since correlated uncertainties would restrict the available parameter space.
\end{description}

In summary, we have $21$ constraints from Lattice QCD calculations, $9$ constraints from SCET relations, and $6$ non-linear constraints from the dispersive bounds.
We test three fit scenarios, by fixing in \eq{eq:z_exp} $N=\{1,2,3\}$.
These three scenarios have $17$, $31$ and $45$ free parameters, respectively, which implies that the $N=\{2,3\}$ cases are under-constrained.
However, the presence of the dispersive bounds ensures bounded posterior distributions despite the appearance of blind directions.

\begin{table}[t]
    \centering
    \begin{tabular}{c c}
    \toprule
        $\mLb$ & $5.620\, \mathrm{GeV}$  \\
        $\mLst$ & $1.520\, \mathrm{GeV}$  \\
        $m_B$ & $5.279\, \mathrm{GeV}$  \\
        $m_K$ & $0.494\, \mathrm{GeV}$  \\
        $m_{K^*}$ & $0.896\, \mathrm{GeV}$  \\
        $\alpha_\mathrm{EM}$ & $1/133$ \\
        $\tau_{\Lb}$ & 1.471 ps\\
    \bottomrule
    \end{tabular}
    \caption{Numerical inputs of our parametrisation. All inputs are taken from the PDG \cite{ParticleDataGroup:2020ssz}, apart from the $\Lambda_b$ lifetime \cite{Amhis:2020phx}.}
    \label{tab:constants}
\end{table}

All the numerical studies are performed using the version 1.0.4 of the \texttt{EOS} software \cite{EOS:paper,EOS:repo}.
We use \texttt{EOS} default values for the physical constants, which include the values in \Table{tab:constants}.
Posterior samples are drawn using a Markov chain Monte-Carlo method based on the Metropolis-Hastings algorithm \cite{Metropolis:1953am, Hastings:1970aa}.
Since the posterior-predictive samples for the parameters of the $z$ expansion are usually non-Gaussian distributed, we provide an ancillary notebook that can be used to reproduce them.
Running the notebook requires only an installation of \texttt{EOS} v1.0.4 or greater\footnote{See \url{https://eos.github.io/doc/installation.html} for installation instructions.}.

\begin{figure}
\hspace{-10mm}
    \begin{tabular}{ccc}
        \includegraphics[width=.35\textwidth]{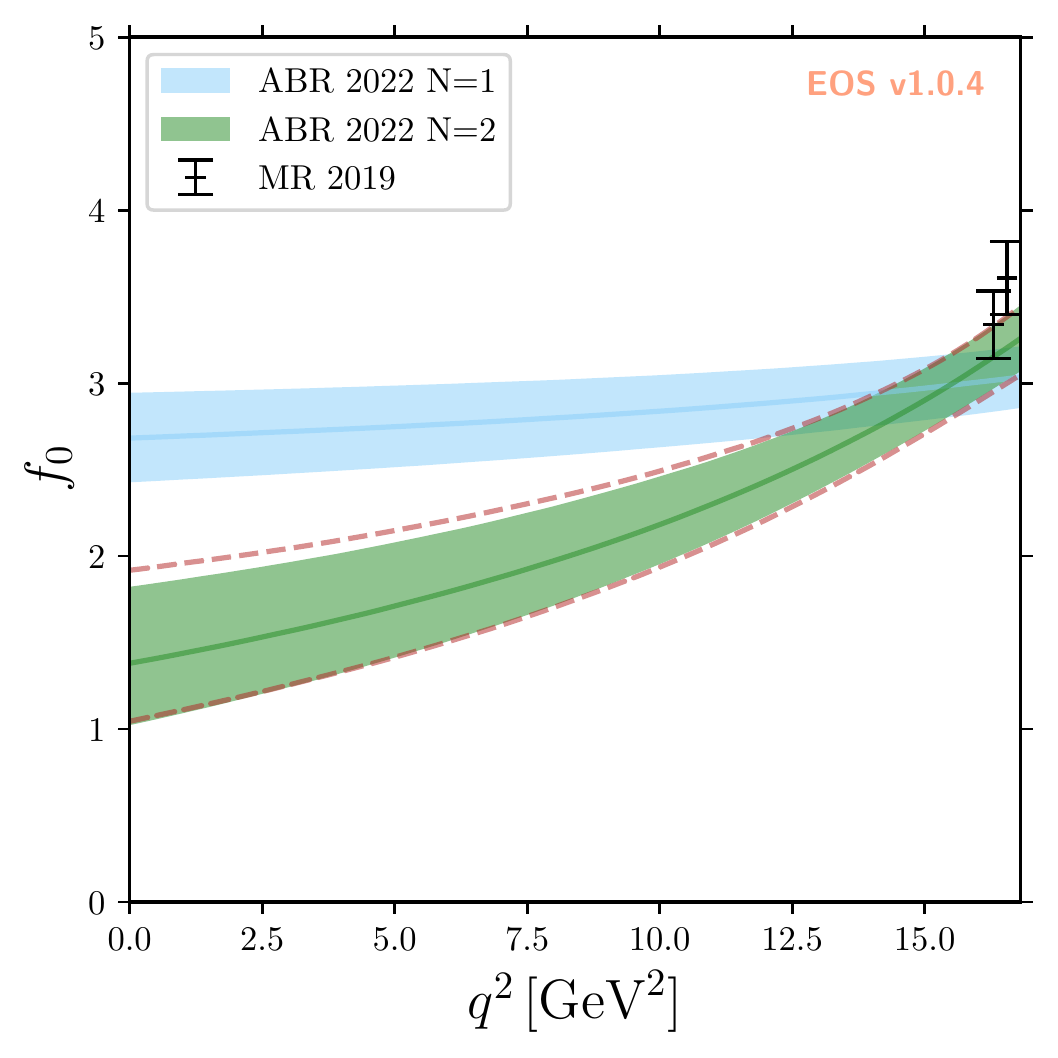}  &
        \includegraphics[width=.35\textwidth]{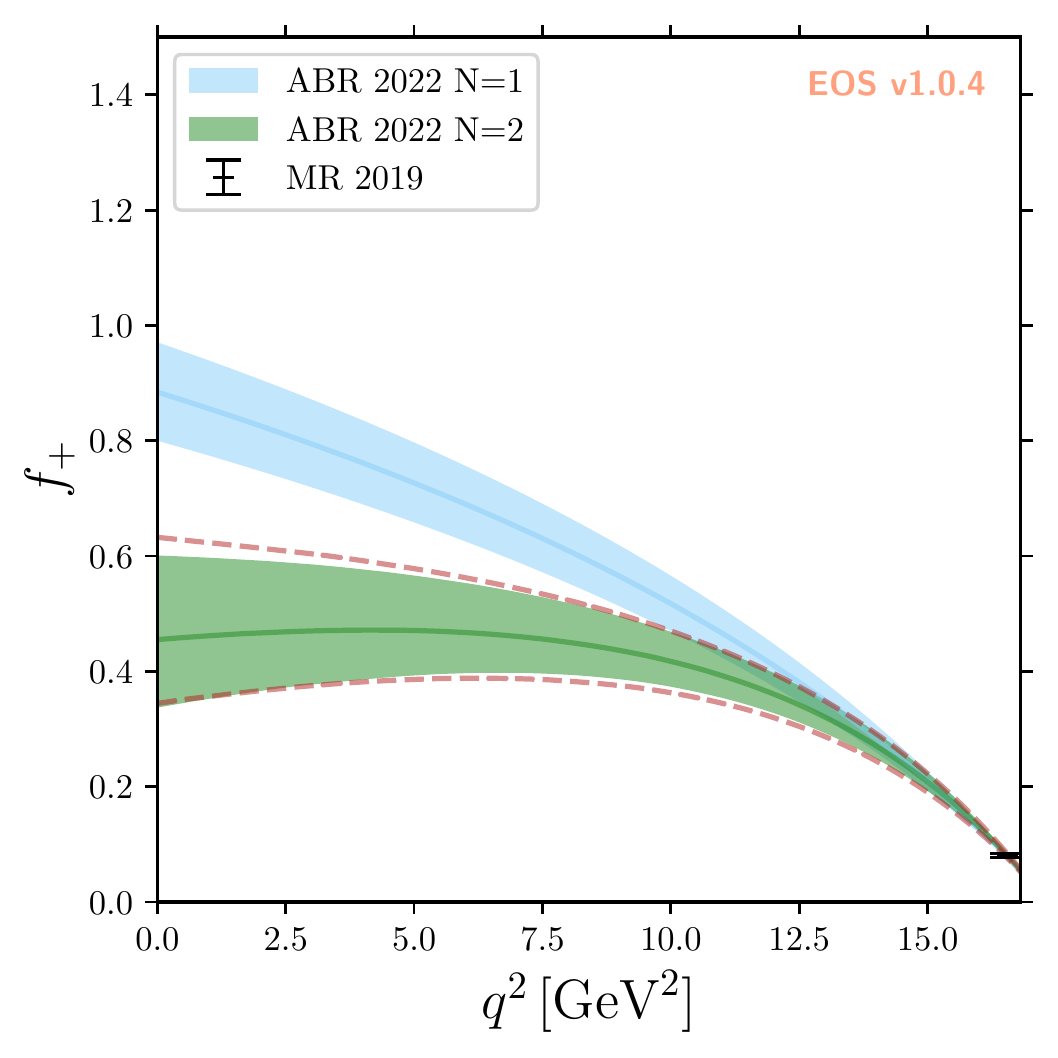} &
        \includegraphics[width=.35\textwidth]{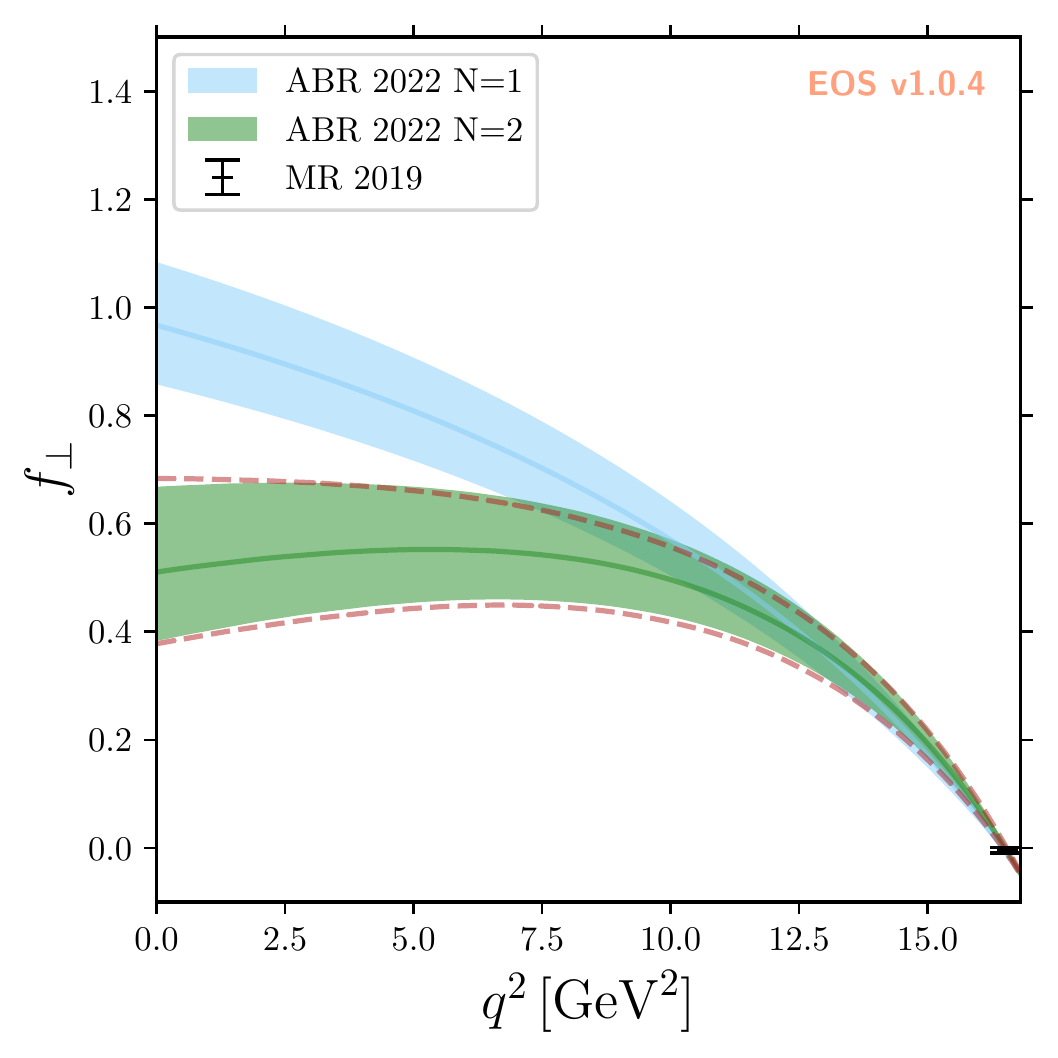}
        \\
        \includegraphics[width=.35\textwidth]{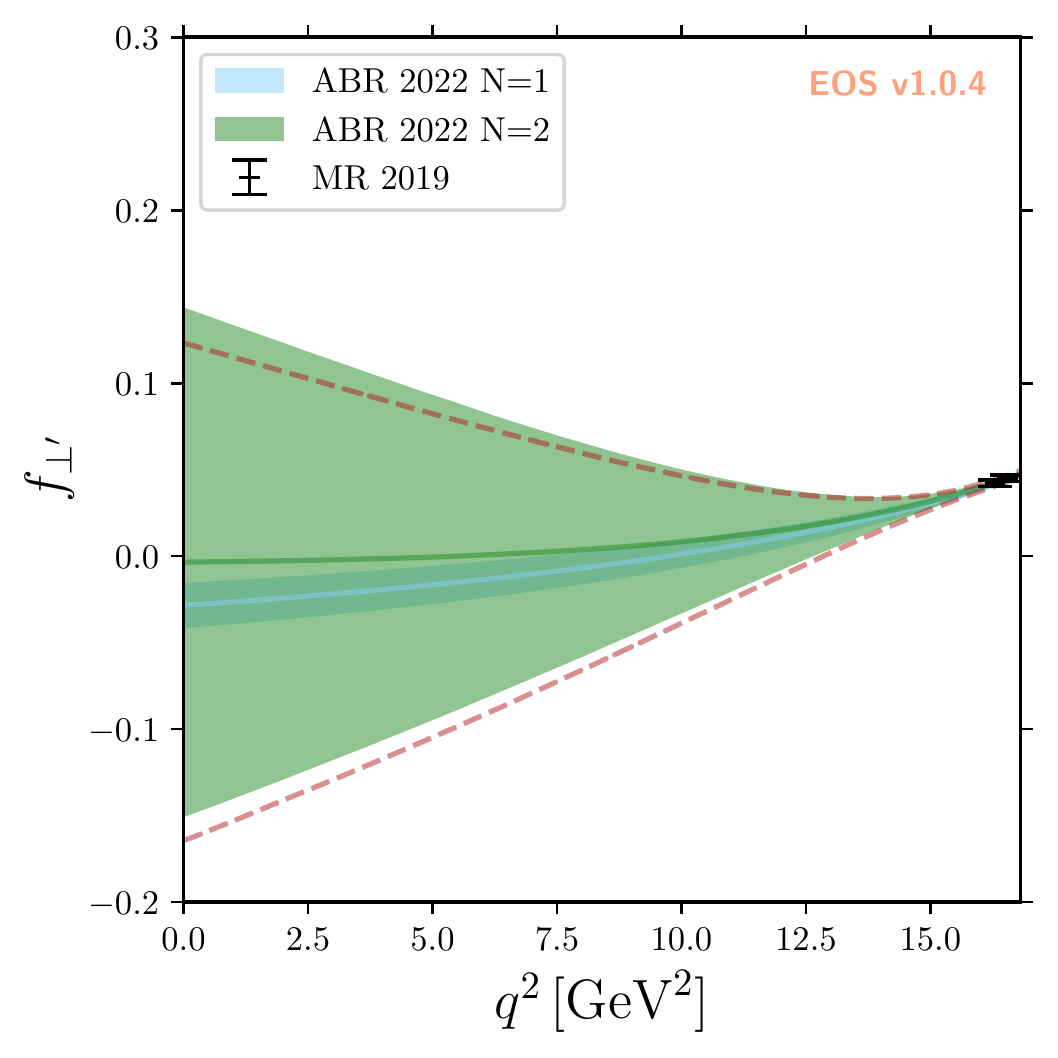}&
        \includegraphics[width=.35\textwidth]{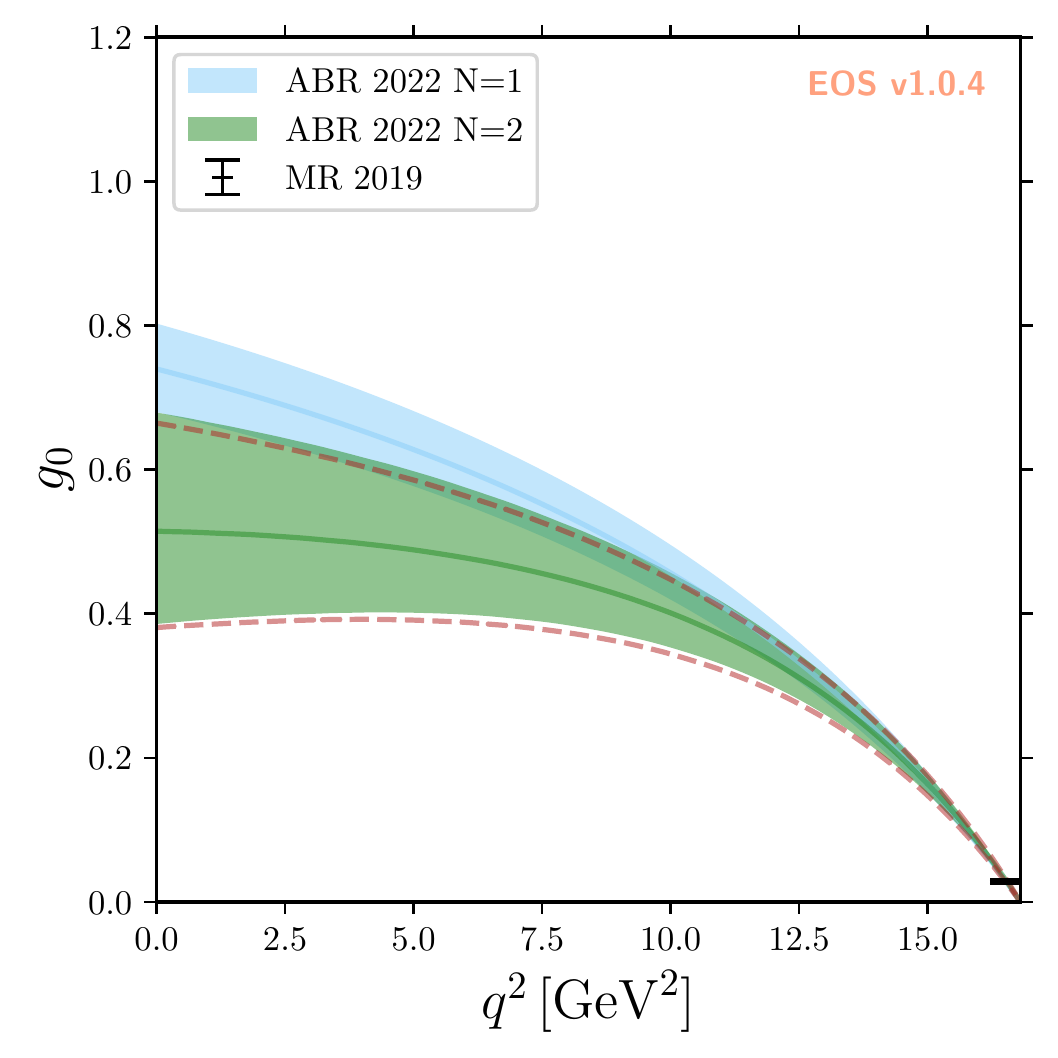}  &
        \includegraphics[width=.35\textwidth]{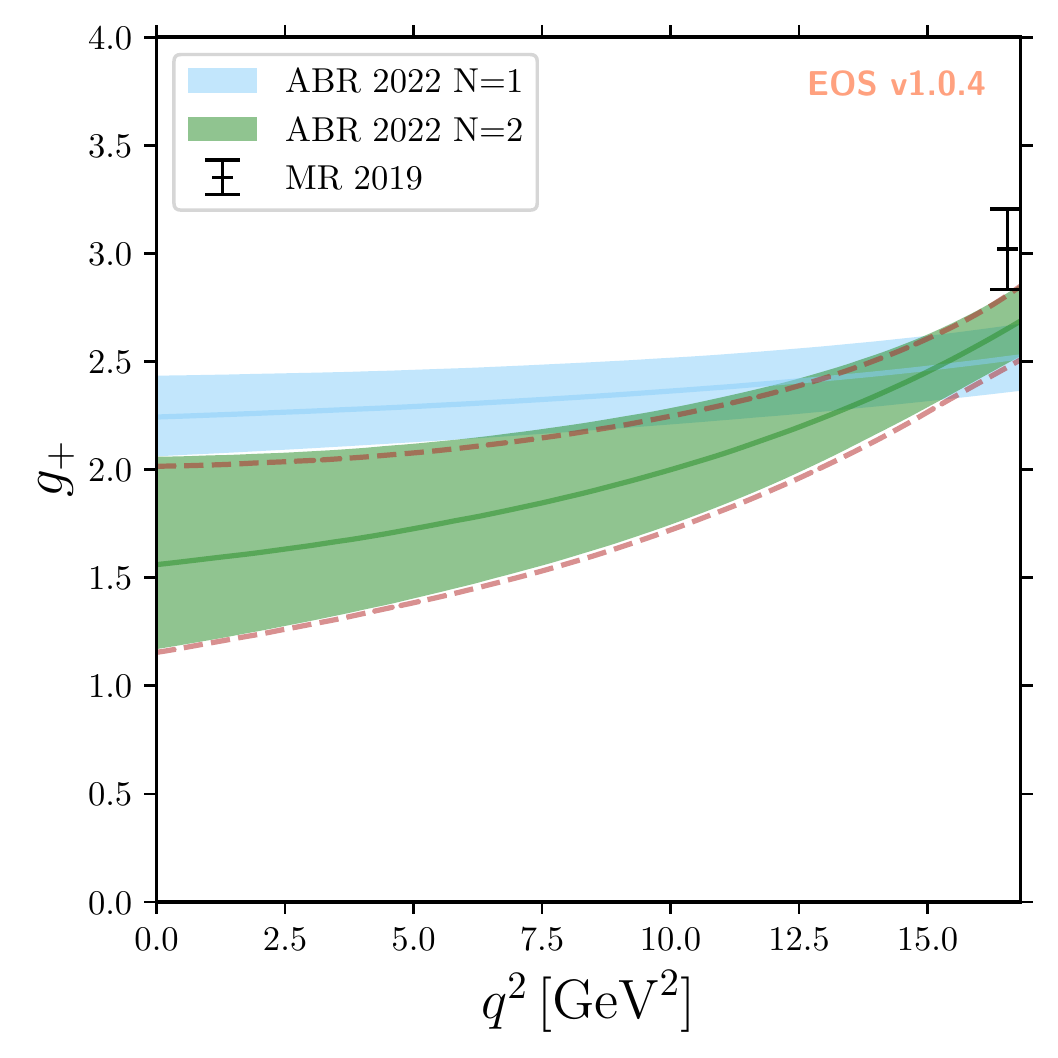}  \\[1.25em]
        \includegraphics[width=.35\textwidth]{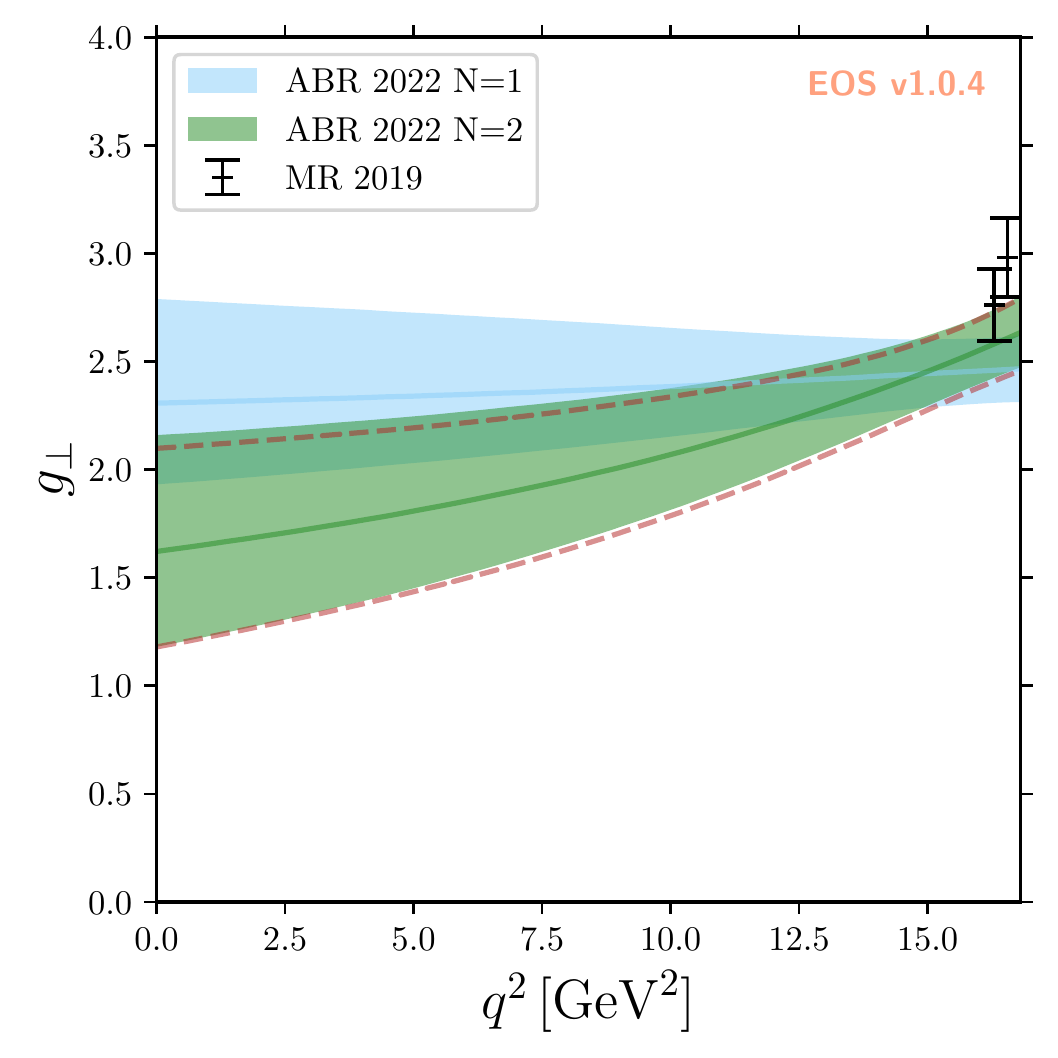} &
        \includegraphics[width=.35\textwidth]{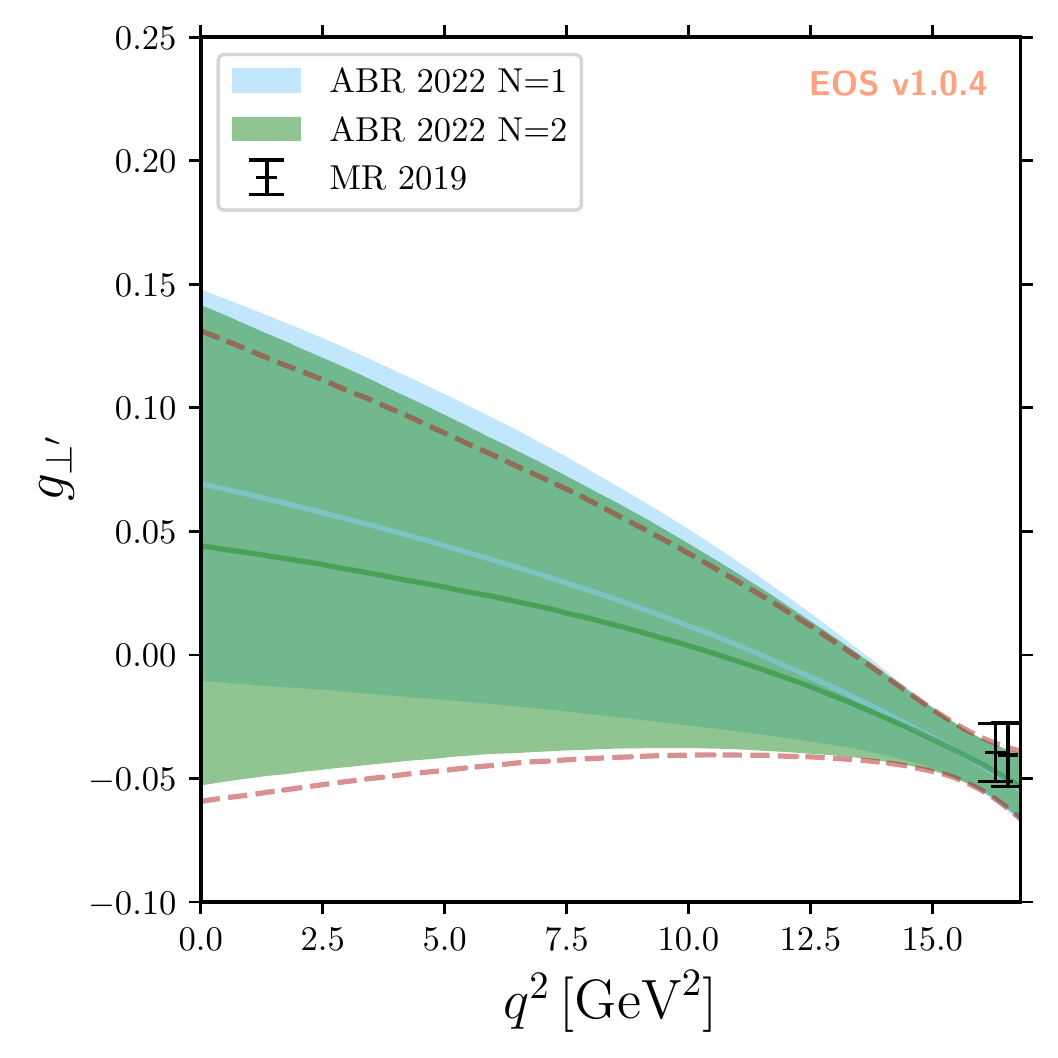}\\[1.25em]
    \end{tabular}
    \caption{Fit results for the vector and axial-vector $\Lambda_b\to\Lambda(1520)$ form factors. The blue and green bands are the 68\% C.L. region, and the solid line shows the median of the distribution for $N=1$ and $N=2$, respectively. The dashed red lines contain the 68\% C.L. region for the $N=3$ scenario.
    }
    \label{fig:fitres1}
\end{figure}

In the following, we discuss our findings.
The results are summarized in Table \ref{tab:fit_summary} and drawn in \figs{fig:fitres1}{fig:fitres2}.
The blue and green bands are obtained by fixing in \eq{eq:z_exp} $N=\{1,2\}$ and we added the $N=3$ scenario in red dashed lines.
These fit scenarios have different characteristics, which we describe below.

\begin{table}[t]
    \centering
    \begin{tabular}{lccccc}
    \toprule
        \multirow{2}{*}{$N$} & \multirow{2}{*}{parameters} & \multicolumn{4}{c}{$-2 \log \mathcal{L}$} \\
        & & Lattice & SCET & Bounds & Global \\
    \midrule
        1 & 17 & 48.4 & 2.6 & 0.1 & 51.1\\
        2 & 31 & 23.6 & 0.3 & 0.1 & 24.0\\
        3 & 45 & 19.6 & 0.2 & 0.1 & 19.9\\
    \bottomrule
    \end{tabular}
    \caption{Summary of the fit results for the different scenarios.
    The number of constraints is the same for all the scenarios, namely 30 linear constraints and 6 non-linear ones.
    Since the bounds provide non-linear constraints on the parameters, the corresponding test statistic is not known.
    For the SCET relations and the lattice data, the contribution to the likelihood $\mathcal{L}$ can be interpreted as a $\chi^2$.}
    \label{tab:fit_summary}
\end{table}

\begin{description}
\item[The $N=1$ scenario.] This parametrisation resembles the one used for the continuum extrapolation of the lattice QCD results in that we have the same number of free parameters. In fact, a fit to the lattice QCD data only, without endpoint relations at low $q^2$, yields a $\chi^2=0$ at the best-fit point. However, when adding endpoint relations at low $q^2$, SCET relations and dispersive bounds, the fit quality drops to a $p$-value of $10^{-6}$. This is mainly due to the incompatibility of lattice QCD data and endpoint relations at low $q^2$ in this scenario. This is expected since lattice QCD data are only available at very high $q^2$. 
It is also evident from \figs{fig:fitres1}{fig:fitres2} that the $N=1$ scenario massively underestimates the uncertainties at large recoil for some of the form factors. 
Concerning the saturations of the two-particle contributions to the dispersive bounds, they all peak between 5 and 10\%, except the one associated with the pseudo-tensor current which saturates the inequality.

\item[The $N=2$ scenario.] This scenario is characterised by a number of free parameters larger than the number of constraints.
Hence, we are unable to provide a global value for the fit quality. The fit to lattice QCD data and the full set of endpoint relations, yields $\chi^2=0$ at the best-fit point. However, the latter violates the dispersive bounds massively.
Performing then the fit including all constraints, namely adding SCET relations and the dispersive bounds, we obtain in the minimum $\chi^2 = 24$. However, this cannot be interpreted with a $p$-value, since we have a negative number of degrees of freedom. It is anyway useful to mention that the $\chi^2$ at the best-fit point associated with the constraints from lattice QCD drops from $\sim 48.4$ in the $N=1$ scenario to $23.6$ in the $N=2$ case.
Since the number of constraints due to the lattice data is 21, this is equivalent to a satisfactory, \emph{individual} $p$-value of $31\%$ for the lattice points only for $N=2$.
Since the other constraints are equally satisfied, we conclude that the $N=2$ scenario shows an overall better agreement with the data.
In this case, having more free parameters than constraints makes the dispersive bound essential and, therefore, leads to larger relative saturations.

\item[The $N=3$ scenario.] This model has even a larger number of free parameters, allowing a perfect fit to lattice QCD data, the full set of endpoint relations and SCET relations. Also in this case, the fit results, however, massively violate the dispersive bounds.
Adding the latter as constraints, we obtain a \emph{individual} $p$-value for the lattice points of $55\%$.
As visible in \figs{fig:fitres1}{fig:fitres2}, the 68\% C.L. bands overlaps perfectly with the one for $N=2$ results.
This means that the uncertainties are saturated due to the fact that the free parameters are all constrained by the dispersive bounds.
We expect this property to be equally verified by $N>3$ scenarios. Concerning the saturations, similar conclusions as for the $N=2$ case apply.
\end{description}

\begin{figure}[H]
\hspace{-10mm}
    \begin{tabular}{ccc}
        \includegraphics[width=.35\textwidth]{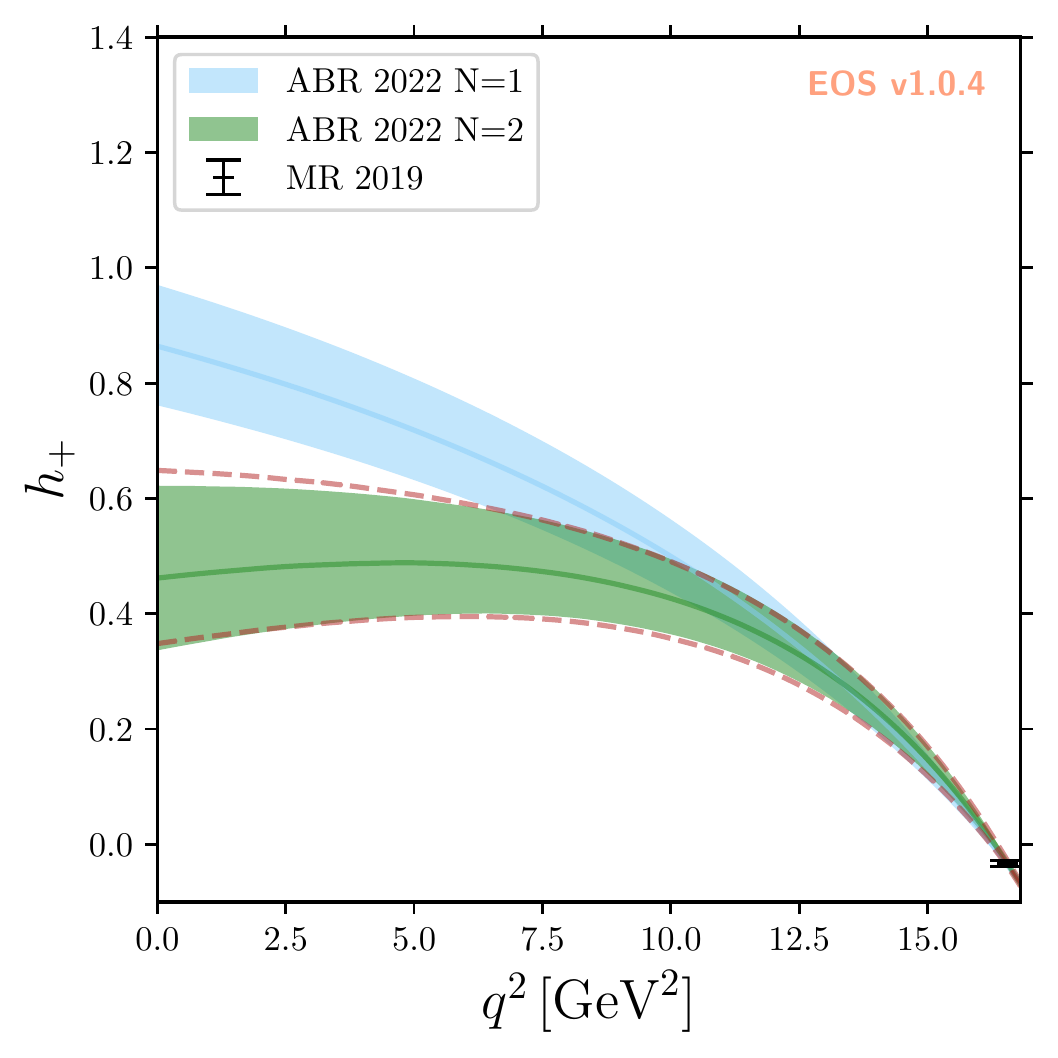} &
        \includegraphics[width=.35\textwidth]{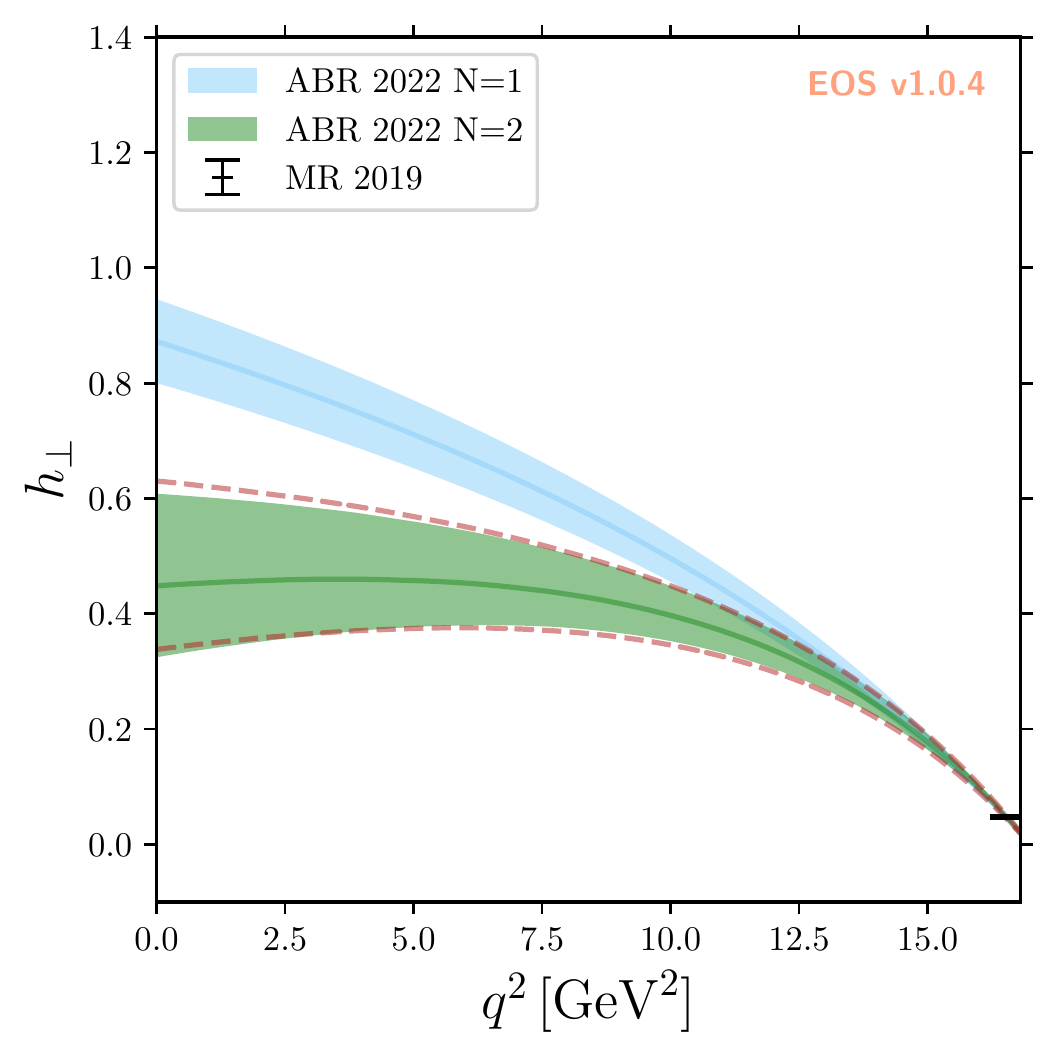} &
        \includegraphics[width=.35\textwidth]{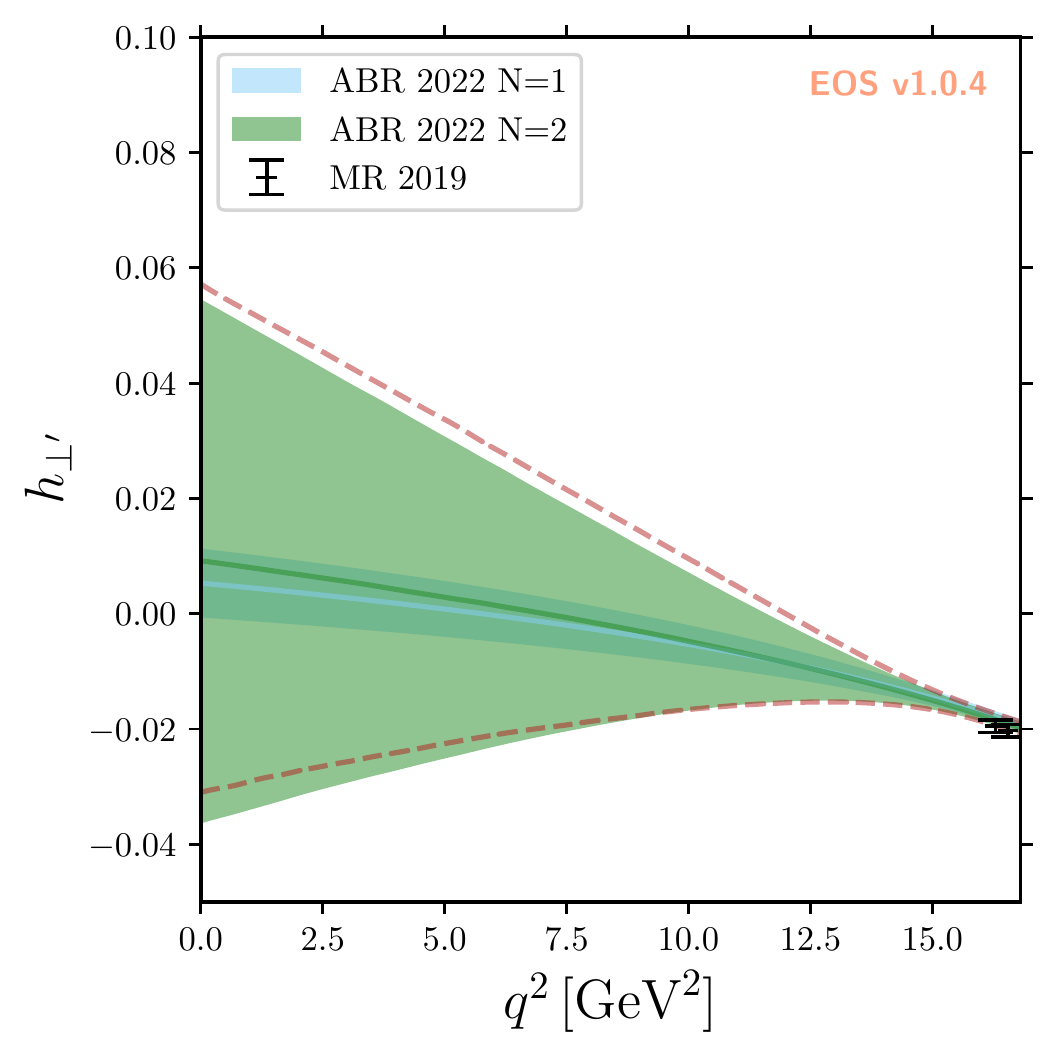}
        \\[1.25em]
        \includegraphics[width=.35\textwidth]{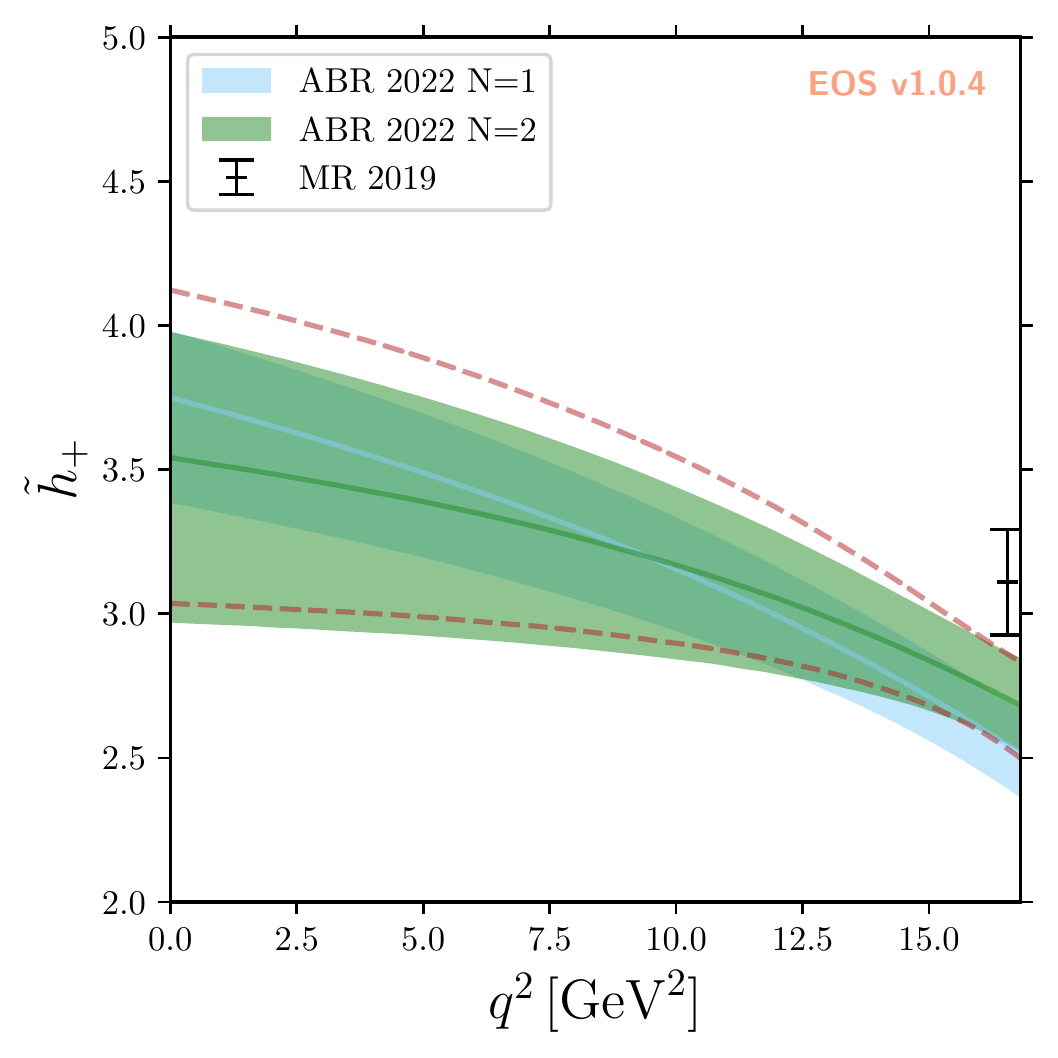} &
        \includegraphics[width=.35\textwidth]{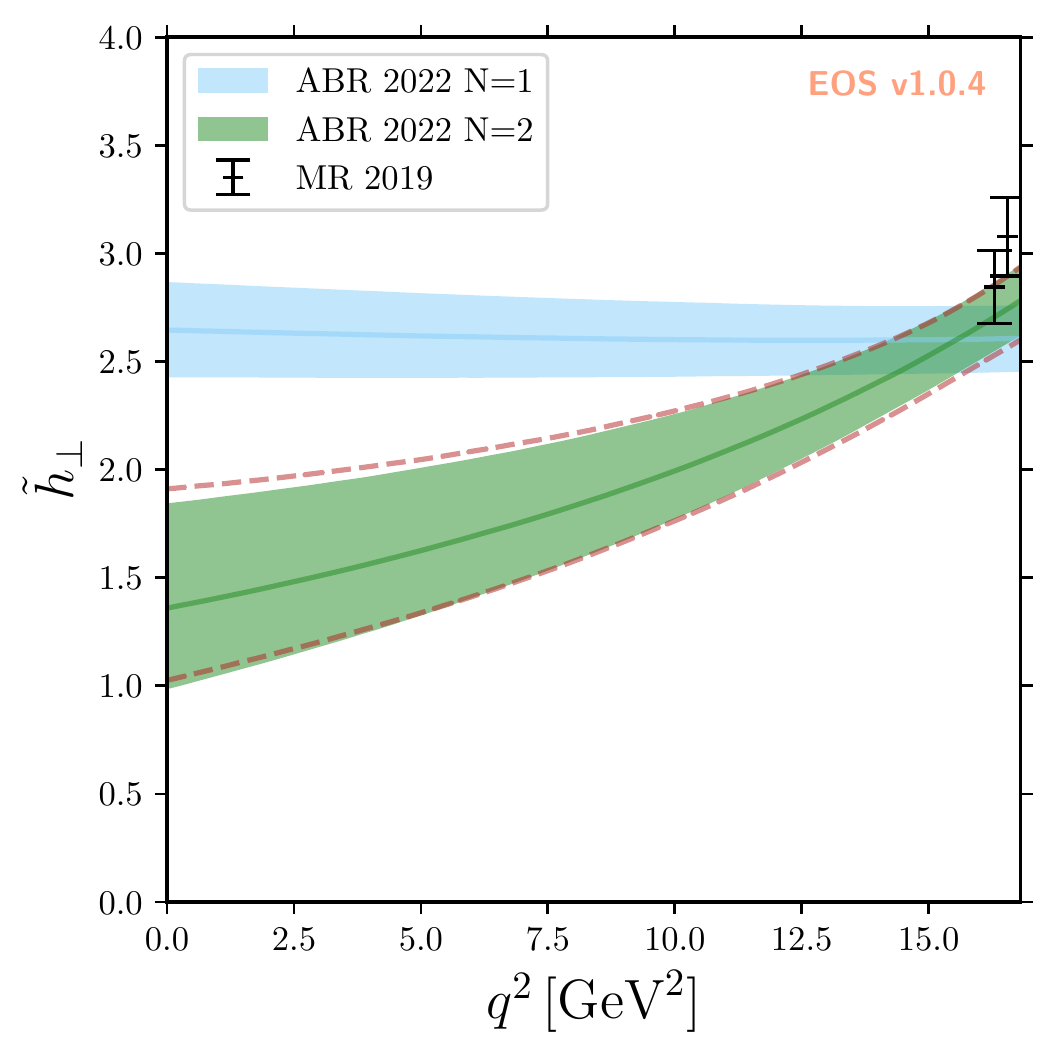} &
        \includegraphics[width=.35\textwidth]{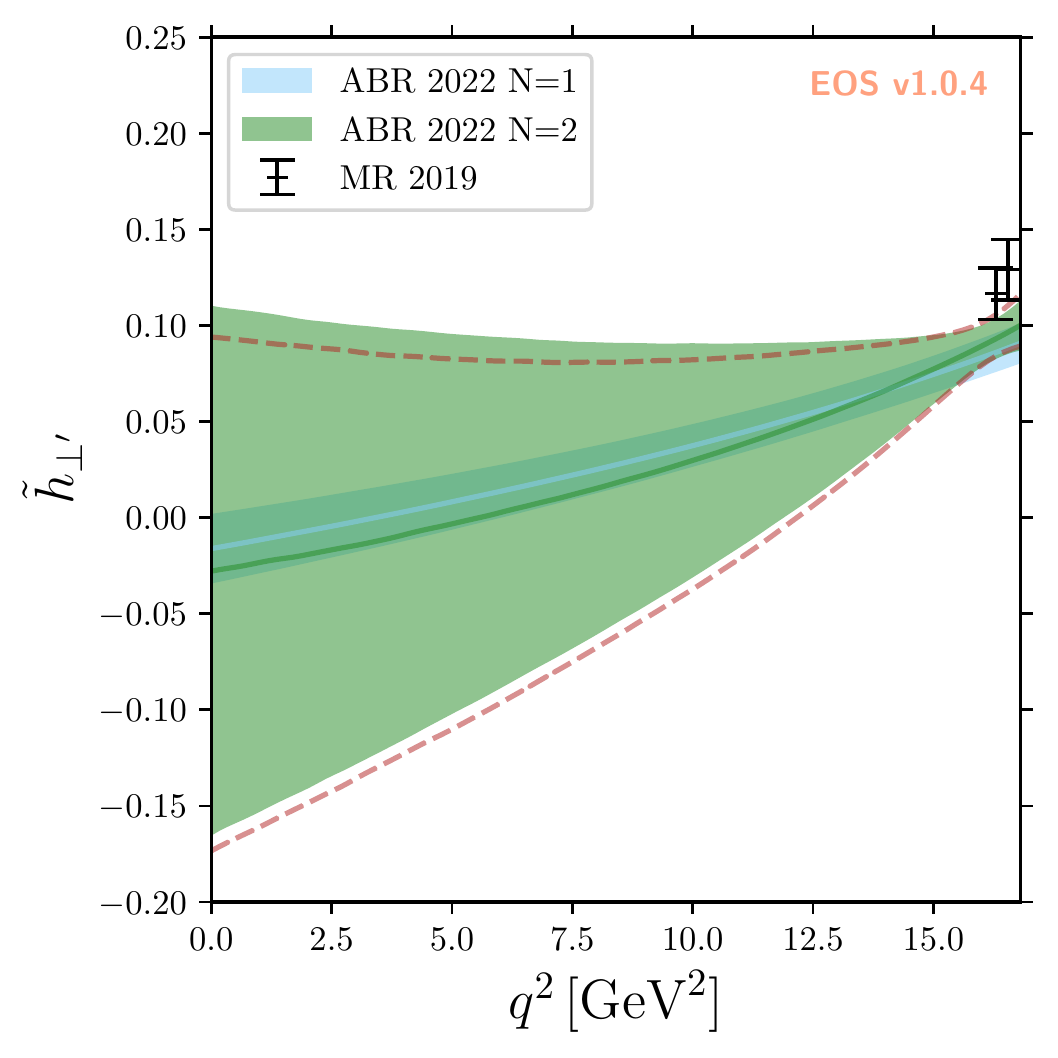}
        \\[1.25em]
    \end{tabular}
    \caption{Fit results for the tensor and pseudo-tensor $\Lambda_b\to\Lambda(1520)$ form factors. The blue and green bands are the 68\% C.L. region, and the solid line shows the median of the distribution for $N=1$ and $N=2$, respectively. The dashed red lines contain the 68\% C.L. region for the $N=3$ scenario.
    }
    \label{fig:fitres2}
\end{figure}

Comparing our findings, we conclude that none of the scenarios is fully satisfactory, because a tension between lattice QCD data and the dispersive bounds remains.
Our results clearly indicate that $N=1$ scenario is ill-suited for an extrapolation of the lattice QCD results to small $q^2$.
For $N>1$ however, the negative number of degrees of freedom ensures the uncertainties to be completely driven by the dispersive bounds.
The latter, therefore, already accounts for the truncation error of the parametrisation and is resistant to any residual model uncertainty.
Despite the residual tension, we can thus set $N=2$ as our nominal fit for the phenomenology applications and, with the current data, considering higher extrapolation orders will not modify our results nor the attached uncertainties.

We emphasize that our approach is systematically improvable by computations of the form factors in the large-recoil region.
For example, future lattice simulations that extend to lower $q^2$ values will provide means to extract more precise information on the higher order coefficients, possibly also removing the aforementioned current tension.
Since the $N=2,\,3$ scenarios show a saturation of the dispersive bound, a global analysis of all $b\to s$ transitions would help reduce the uncertainties on all the form factors.

\section{Phenomenology}
\label{sec:pheno}
The effective Hamiltonian describing $b\to s\ell^+\ell^-$ transitions at the scale $\mu=m_b$ reads \cite{Buchalla:1995vs}:
\begin{equation}
    \mathcal{H}(b\to s\ell^+\ell^-)=-\frac{4G_F}{\sqrt{2}}V_{tb}V^*_{ts}\sum_{i=1}^{10} C_i\mathcal{O}_i\,,
    \label{eq:Heffbsll}
\end{equation}
where the relevant operators for our discussion are the four quarks operators:
\begin{align}
    \mathcal{O}_1 =& (\bar{c}\gamma_\mu T^a P_L c)(\bar{s} \gamma_\mu T^a P_{L} b) & \mathcal{O}_2 =& (\bar{c}\gamma_\mu P_L c)(\bar{s} \gamma_\mu P_{L} b)\,,
\end{align}
with $C_1 =1.010$ and $C_2=-0.291$ \cite{Gorbahn:2004my,Bobeth:1999ww}, and the semileptonic and dipole operators:
\begin{align}
    \mathcal{O}_9^{(\prime)} =&\, \frac{e^2}{16\pi^2}(\bar{s} \gamma_\mu P_{L(R)} b)(\bar{\ell}\gamma^\mu\ell)\,, &  \mathcal{O}_{10}^{(\prime)} =&\, \frac{e^2}{16\pi^2}(\bar{s} \gamma_\mu P_{L(R)} b)(\bar{\ell}\gamma^\mu\gamma_5\ell)\,, \\[0.5em]
    \mathcal{O}^{(\prime)}_7=&\, \frac{e\,m_b}{16\pi^2}(\bar{s}\sigma_{\mu\nu}P_{R(L)} b)F^{\mu\nu}\,,
\end{align}
where in the SM $C_9\simeq 4.2$, $C_9\simeq -4.3$, $C_7\simeq-0.29$, and $C_7^\prime \sim m_s/m_b \approx 0$, while $C_9^\prime = C_{10}^\prime =0$ \cite{Gorbahn:2004my,Bobeth:1999ww}.
The operators $\mathcal{O}_{1-6}$ mix perturbatively onto $\mathcal{O}_9$ and $\mathcal{O}_7$, even though $C_{3-6}\sim \mathcal{O}(10^{-2})$ only generate small effects.
The mixing has been calculated at leading logarithmic power, and is accounted for by the shift $C_9\to C_9^\mathrm{eff}(q^2)$ and $C_7\to C_7^\mathrm{eff}(q^2)$ \cite{Buchalla:1995vs,Chetyrkin:1996vx}. 
The operators $\mathcal{O}_{1,2}$ are also responsible for non-factorisable long distance contributions in $b\to s\ell^+\ell^-$ transitions.
In the literature, these long-distance effects have been studied in different setups \cite{Gubernari:2022hxn,Capdevila:2017ert,Khodjamirian:2010vf,Bobeth:2017vxj}.
A detailed analysis of these effects is beyond the scope of this work.
Therefore, we restrain ourselves from providing any prediction where non-factorisable long-distance effects dominate, namely the $c\bar{c}$ resonant region.\\
The effective Hamiltonian in \eq{eq:Heffbsll} is used to derive several observables for $\Lb\to\Lst\ell^+\ell^-$ decays.
We refer to Refs.~\cite{Descotes-Genon:2019dbw, Das:2020cpv} for the full expressions of the 4-fold differential distribution for $\Lb\to\Lst(\to N\bar{K})\ell^+\ell^-$ with $N=p,n$.
Note that Ref.~\cite{Descotes-Genon:2019dbw} neglects lepton masses, which is not the case here.
In the following, we always integrate on the $\Lst\to N\bar{K}$ variables, in the hypothesis of a narrow $\Lst$, such that the remaining distribution for $\Lambda_b\to\Lst\ell^+\ell^-$ decays is
\begin{equation}
    \frac{d\Gamma}{dq^2 d\cos\theta_\ell} = a + b \cos\theta_\ell + c \cos^2\theta_\ell\,,
\end{equation}
where $\theta_\ell$ is the helicity angle of the negative-charged lepton in the dilepton rest-frame, and 
\begin{equation}
\begin{aligned}
    a = &\; \frac{1}{2}(L_{1ss}+2 L_{2ss}+L_{3ss})\,, & b = &\; \frac{1}{2}(L_{1c}+2L_{2c}) \,,\\
    c = &\; \frac{1}{2}(L_{1cc}-L_{1ss}+2L_{2cc}-2L_{2ss}-L_{3ss})  \,,
\end{aligned}
\end{equation}
and the coefficients $L_i$ are listed in Ref.~\cite{Descotes-Genon:2019dbw}.
With these, we define the following differential and integrated observables:
\begin{equation}
\begin{aligned}
    \frac{d\mathcal{B}(q^2)}{dq^2}=&\,2\,\tau_{\Lambda_b}\left(a+\frac{c}{3}\right)\,,   &  
    \mathcal{B}[q_\mathrm{min}^2,q_\mathrm{max}^2]=&\,2\tau_{\Lambda_b}\int_{q^2_\mathrm{min}}^{q^2_\mathrm{max}}dq^2\left(a+\frac{c}{3}\right)\,,  \\ \mathcal{A}^\ell_\mathrm{FB}(q^2) =&\,\frac{b}{2\left(a+\frac{c}{3}\right)}\,, 
     &  \mathcal{A}^\ell_\mathrm{FB}[q_\mathrm{min}^2,q_\mathrm{max}^2] =&\,\frac{\int_{q^2_\mathrm{min}}^{q^2_\mathrm{max}}dq^2 b}{2\int_{q^2_\mathrm{min}}^{q^2_\mathrm{max}}dq^2\left(a+\frac{c}{3}\right)} \,, \\
   S_{1cc}(q^2)=&\,\frac{L_{1cc}+\bar{L}_{1cc}}{\frac{d\Gamma}{dq^2}+\frac{d\bar\Gamma}{dq^2}}\,,  &  S_{1cc}[q^2_\mathrm{min},q^2_\mathrm{max}]=&\,\frac{\int_{q^2_\mathrm{min}}^{q^2_\mathrm{max}}dq^2\left(L_{1cc}+\bar{L}_{1cc}\right)}{\int_{q^2_\mathrm{min}}^{q^2_\mathrm{max}} dq^2 \left(\frac{d\Gamma}{dq^2}+\frac{d\bar\Gamma}{dq^2}\right)}\,,
\end{aligned}
\label{eq:observables}
\end{equation}
where the bar denotes CP-conjugated quantities, and
\begin{equation}
    \frac{d\Gamma}{dq^2} = \frac{1}{\tau_{\Lambda_b}}\frac{d\mathcal{B}(q^2)}{dq^2}
\end{equation}
The observables above are chosen because they are the most promising ones to be measured at LHCb \cite{Amhis:2020phx}.\\

\begin{figure}[h]
    \centering
    \includegraphics[width=.4\textwidth]{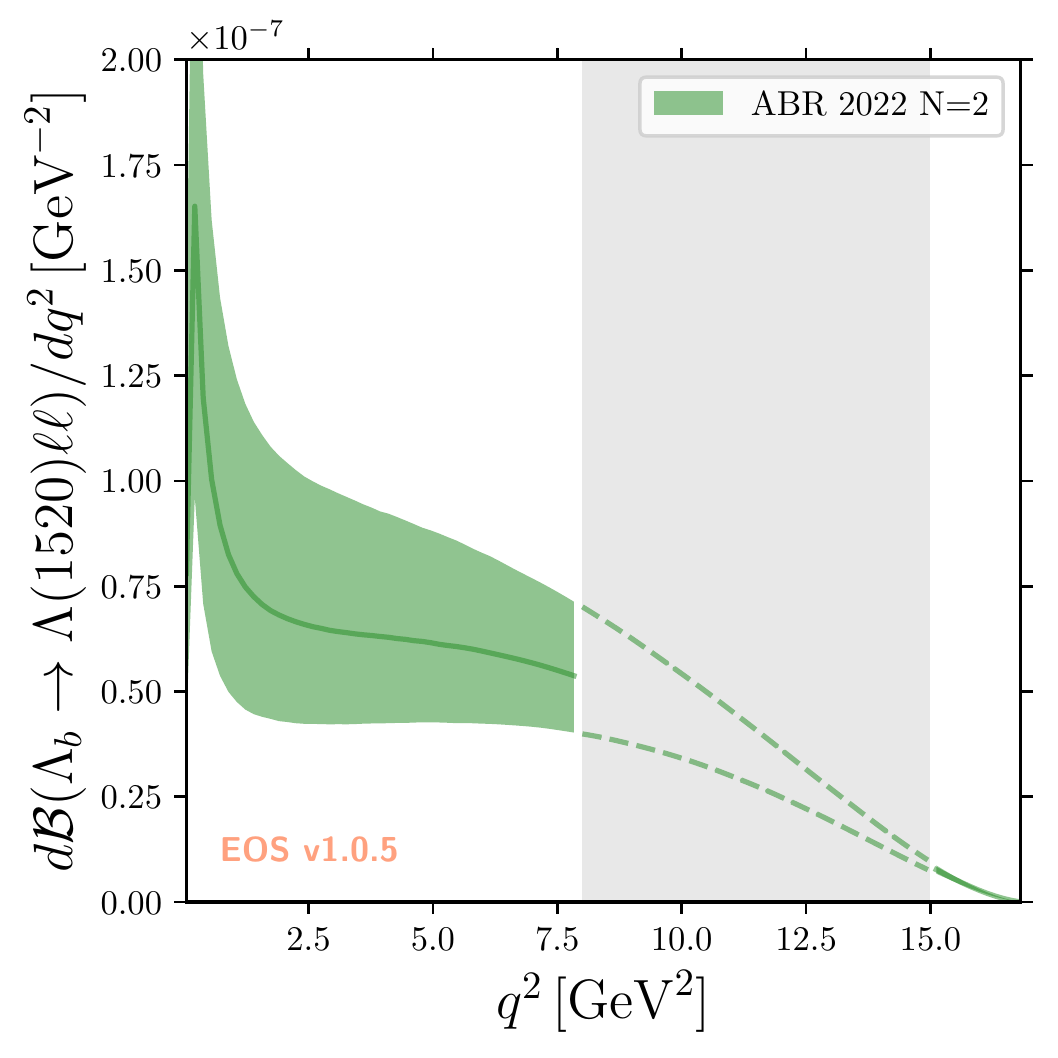}
    \includegraphics[width=.4\textwidth]{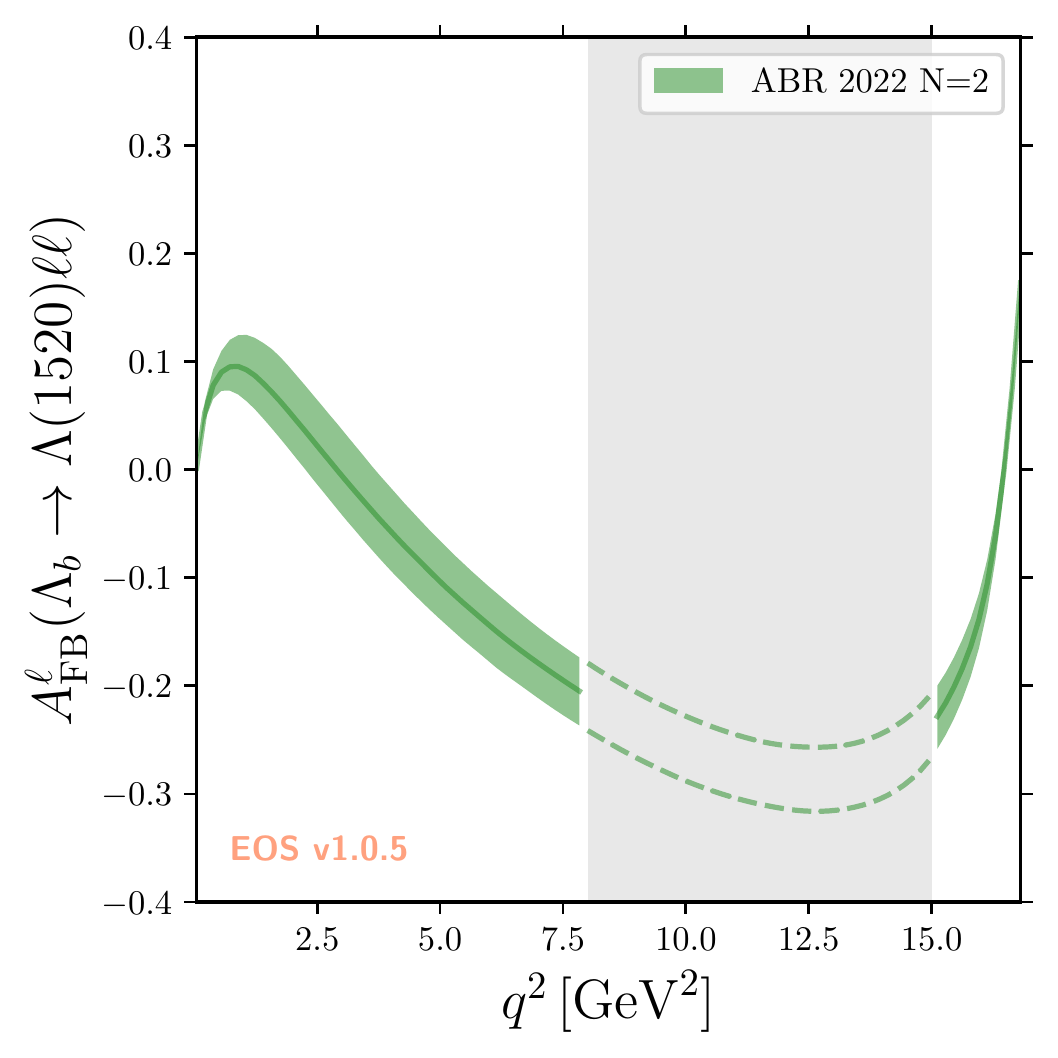} \\
    \includegraphics[width=.4\textwidth]{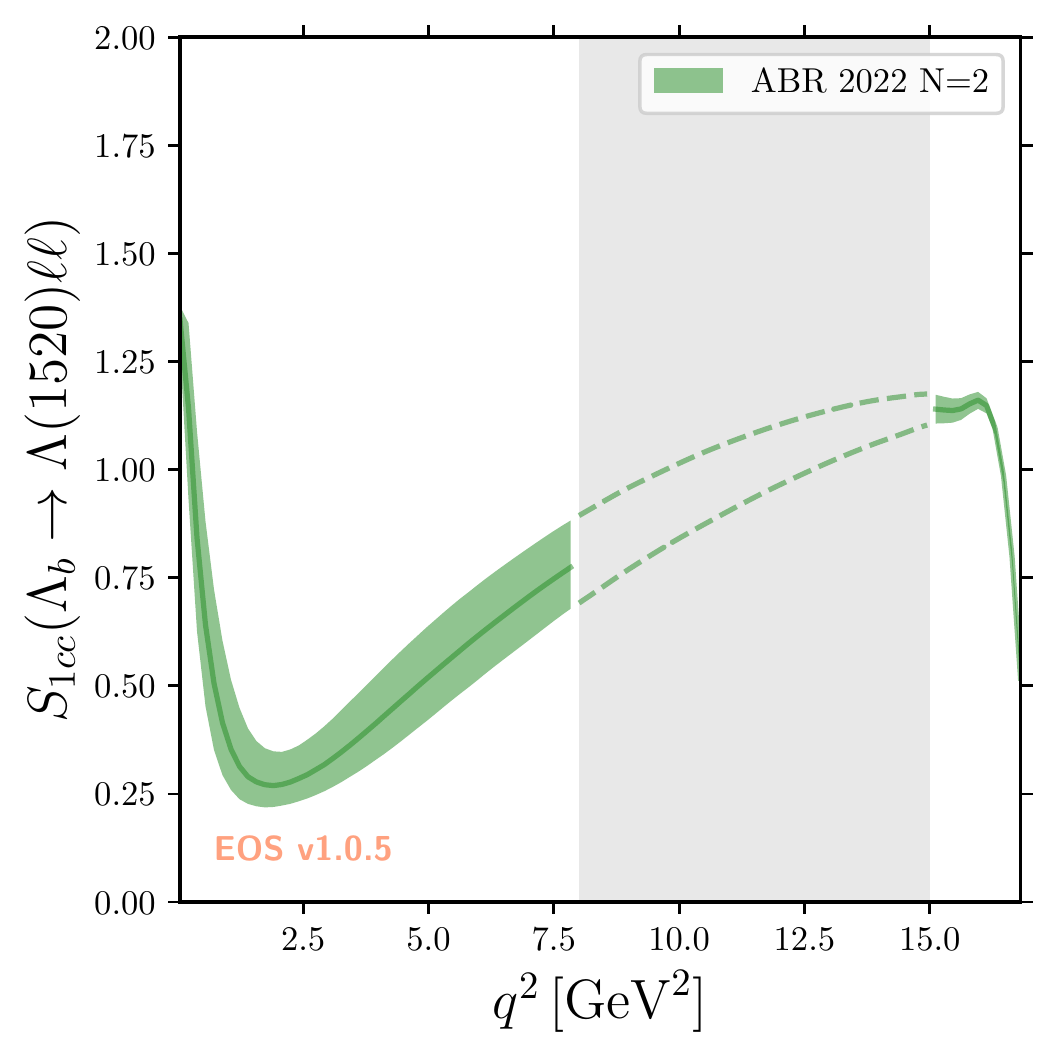}
    \caption{Plots of the differential branching fraction (upper left plot), differential forward-backwards lepton asymmetry (upper right plot) and differential $S_{1cc}$ (lower plot). The green bands are the 68\% C.L. region, and the full line is the median of the distribution in the nominal fit. We use dashed lines in the vetoed resonant region.
    }
    \label{fig:phenores}
\end{figure}

We use now the results in \sec{sec:fit} to predict the above observables. We stress that we use as nominal fit results for the form factors parameters the $N=2$ scenario. 
We plot differential distributions in \fig{fig:phenores}.
As already mentioned, we veto the $c\bar{c}$ resonant region ($8\,\mathrm{GeV}^2<q^2<15\,\mathrm{GeV}^2$), but we show for completeness the naive extrapolation in the resonances region with dashed lines. \\
In the left upper panel of \fig{fig:phenores}, we show the differential branching ratio. We compare our results with the literature \cite{Descotes-Genon:2019dbw,Mott:2011cx,Das:2020cpv} and find some discrepancies.
We notice that the order of magnitude of our prediction of the differential branching ratio differs, in the entire $q^2$ range, from the one quoted in Ref.\cite{Descotes-Genon:2019dbw}. 
We then compare with the most recent measurement of $\mathcal{B}(\Lb\to p \bar{K} \mu^+\mu^-)$ in Ref.\cite{LHCb:2019efc}, finding a good agreement between the order of magnitude of the latter and our predictions, even though the various $\Lambda^* \to p \bar{K}$ resonances are not separated.
Furthermore, we also see a difference in the branching ratio as a function of $q^2$ at very large recoil. We trace back this difference to the different form factor parametrisation.
We verified explicitly that the Quark Model predictions in their current form violate SCET relations at low $q^2$ as well as the dispersive bounds, even of more than an order of magnitude for some currents.
Without a reasonable estimation of the uncertainties and correlations of the Quark Model parameters, it is impossible to extract more meaningful information.
Also, for some of the form factors, lattice QCD results and Quark Model ones are not in good agreement close to zero-recoil.\\
Finally, we observe that the lack of theory points for the form factors at large recoil results in large uncertainties.
We stress that due to this, the shape of the differential branching ratio in the central $q^2$ region could be qualitatively different when new lattice data or QCD Sum Rule-based computation will be available.
This model dependency is less evident in the differential forward-backwards asymmetry, right panel of \fig{fig:phenores}.
We note that, differently than in the case of $B\to K^*\ell^+\ell^-$ and $\Lambda_b\to\Lambda\ell^+\ell^-$, $\mathcal{A}^\ell_\mathrm{FB}$ for $\Lambda_b\to\Lst\ell^+\ell^-$ shows two zero-crossings, one at low and the other at high-recoil. The one at large recoil has to be understood in the same way as for $B\to K^*\ell^+\ell^-$ and $\Lb\to \L\ell^+\ell^-$, namely due to a combination of $C_7$ and $C_9$ that vanishes for some $q^2$ value. The zero crossing at low recoil is instead of a different nature, since it is due to the zero-crossing in $f_\perp$. This can for example be more easily seen in the Heavy Quark Limit \cite{Descotes-Genon:2019dbw,Amhis:2020phx}, where, at leading order in $\alpha_s$ and leading power in $1/m_b$, $\mathcal{A}^\ell_\mathrm{FB}$ is proportional to $f_\perp$.
We stress that this behaviour is also present in predictions based on the Quark Model \cite{Descotes-Genon:2019dbw,Amhis:2020phx,Mott:2011cx}.
The zero-crossing in $f_\perp$ seems to be a peculiarity of several $1/2^+\to 3/2^-$ transitions \cite{Meinel:2021mdj,Mott:2011cx}, and therefore, any further theory data or experimental measurement of such modes could help to confirm or refute this property. For convenience, we also provide integrated observables in \Table{tab:predictions}.

With our setup we can also predict $\mathcal{B}(\Lb\to\Lst\gamma)$ \cite{Hiller:2007ur}. We have 
\begin{equation}
\begin{aligned}
    \mathcal{B}(\Lb\to\Lst\gamma)=\,&\tau_{\Lambda_b}\frac{G_F^2 |V_{tb}V^*_{ts}|^2 \alpha_\mathrm{em}m_b^2}{384\pi^4 \mLb^3}(\mLb^2-\mLst^2)\\
    \times&\bigg[|C_7+C_7^\prime|^2(\mLb+\mLst)^4(|h_\perp(0)|^2+3|h_{\perp^\prime}(0)|^2 ) \\
    +&|C_7-C_7^\prime|^2(\mLb-\mLst)^4(|\tilde{h}_{\perp}(0)|^2+3|\tilde{h}_{\perp^\prime}(0)|^2)\bigg]\,.
    \end{aligned}
\end{equation}
If we restrict ourselves to the SM, we find:
\begin{equation}
    \mathcal{B}(\Lb\to\Lst\gamma) = (3.5^{+2.7}_{-1.6}) \times 10^{-5} \,.
    \label{eq:pred_LbtoLstgamma}
\end{equation}
Even though the large uncertainties forbid us to obtain a precise prediction, we notice that $\mathcal{B}(\Lb\to\Lst\gamma)$ has the same order of magnitude of LHCb measurement of $\mathcal{B}(\Lb\to\Lambda\gamma)$ \cite{LHCb:2019wwi}. This can be tested by measurements of the $\Lambda_b\to pK\gamma$ resonant spectrum which can be performed at the LHCb experiment\cite{Albrecht:2020azd}.
\begin{table}[h]
    \centering
    \begin{tabular}{c c c }
    \toprule
        $q^2$ bin ($\mathrm{GeV}^2$) & $\mathcal{B}[q^2_\mathrm{min},q^2_\mathrm{max}]\times 10^8$ & $\mathcal{A}^\ell_\mathrm{FB}[q^2_\mathrm{min},q^2_\mathrm{max}] \times 10^{2} $ \\
        \midrule
        $[0.1,3]$   & $24.6^{+13.9}_{ -9.4}$  &  $  6.0^{+2.5}_{-2.5}$ \\[0.5ex]
        $[3,6]$     & $18.7^{ +8.4}_{ -5.9}$  &  $ -7.8^{+3.9}_{-3.5}$ \\[0.5ex]
        $[6,8]$     & $11.3^{ +4.0}_{ -3.0}$  &  $-17.8^{+3.3}_{-3.2}$ \\[0.5ex]
        $[1.1,6]$   & $31.6^{+15.5}_{-10.5}$  &  $ -2.7^{+3.9}_{-3.5}$ \\[0.5ex]
        $[15,q^2_\mathrm{max}]$ & $0.64^{+0.08}_{-0.08}$ & $-18.2^{+2.7}_{-2.7}$ \\
    \bottomrule
    \end{tabular}
    \caption{Predictions of the integrated branching ratio and forward-backwards lepton asymmetry in various $q^2$ bins.}
    \label{tab:predictions}
\end{table}

\section{Conclusions}
\label{sec:conclusions}
In this work, we present a parametrisation of $\L_b\to\L(1520)$ local form factors which respects the analytical properties and the dispersive bounds associated with local $b\to s$ currents. With this setup, we are able to evaluate for the first time dispersive bounds for the form factor coefficients. We determine the form factor parameters through fits to theory data from Lattice QCD, with constraints from endpoint relations, SCET relations, and dispersive bounds. \\
We study the consequences of expanding the form factors up to different orders, finding that the minimal viable scenario is a second-order expansion, and additional orders does not add any information. Hence, we use this fit scenario as the nominal one.
Our results can be easily reproduced using \texttt{EOS} and the python notebook provided as an ancillary file.\\
We provide predictions for differential and integrated branching ratio and leptonic forward-backward asymmetry for $\L_b\to\L(1520)\ell^+\ell^-$ decays, where $\ell$ is a light lepton, and also an update for $\mathcal{B}(\L_b\to\L(1520)\gamma)$.
We stress that our approach suffers from large uncertainties at low $q^2$, due to the lack of calculations of the form factors in that region.
It is, hence, of the utmost importance to focus on that to improve predictions of $\Lb\to\Lambda(1520)$ form factors.

\section*{Acknowledgements}
We thank Claudia Cornella, Carla Marin Benito, Stefan Meinel, Mart\'{i}n Novoa-Brunet, Gumaro Rendon, Danny van Dyk and Felicia Volle for useful discussions.

The work of MR is supported by the DFG within the Emmy Noether Programme under grant DY-130/1-1
and the Sino-German Collaborative Research Center TRR110 ``Symmetries and the Emergence of Structure in QCD''
(DFG Project-ID 196253076, NSFC Grant No. 12070131001, TRR 110). MR is grateful to the CERN Theory Division for the hospitality during parts of this work.

\appendix


\section{Comparison with literature}
In this appendix we provide the relations between the form-factor bases of Refs.~\cite{Meinel:2021mdj} and~\cite{Hiller:2021zth}.
\begin{align}
    f_0 &= \frac{s_+}{\mLst} f_t^{V} & f_+ =& \frac{s_-}{\mLst} f_0^{V} &
    f_\perp =& \frac{s_-}{\mLst} f_\perp^{V} & f_{\perp'} =& f_g^{V} \\
    g_0 &= \frac{s_-}{\mLst} f_t^{A} & g_+ =& \frac{s_+}{\mLst} f_0^{A} &
    g_\perp =& \frac{s_+}{\mLst} f_\perp^{A} & g_{\perp'} =& - f_g^{A}
\end{align}
and
\begin{align}
    h_+ &=  \frac{s_-}{\mLst} f_0^{T} &
    h_\perp =&  \frac{s_-}{\mLst} f_\perp^{T} & h_{\perp'} =& \frac{1}{\mLb+\mLst} f_g^{T} \\
    \tilde{h}_+ &=  \frac{s_+}{\mLst} f_0^{T} &
    \tilde{h}_\perp =& \frac{s_+}{\mLst} f_\perp^{T} & \tilde{h}_{\perp'} =& -\frac{1}{\mLb-\mLst} f_g^{T}
\end{align}

\section{Orthonormal polynomials} 
\label{app:orthonormal_polynomials}
We expand the form factors using the polynomials $p_n$, defined through an orthonormalisation procedure, and referred to as normalized Szeg\H{o} polynomials.
We follow the procedure in Ref.~\cite{Blake:2022vfl}. First of all, we define:
\begin{equation}
    2\, \alpha_{\Lambda_b\Lambda^*} = 2\, \mathrm{Arg}[z((\mLb+\mLst)^2)] = 3.42518
\end{equation}
that determines the arc of the unit circle that we have to integrate on.
The Szeg\H{o} polynomials can be defined through the recurrence relation \cite{Simon2004OrthogonalPO}
\begin{equation}
\begin{aligned}
    \Phi_0(z) &= 1\,,  &  \Phi^*_0(z) &= 1\,, \\
    \Phi_n(z) &= z \, \Phi_{n-1}-\rho_{n-1}\Phi^*_{n-1}\,,  &  \Phi^*_n(z) &= \Phi^*_{n-1}-\rho_{n-1} z \, \Phi_{n-1}\,,
\end{aligned}
\end{equation}
where $\rho_i$ are the so-called Verblunsky coefficients. By imposing orthonormality of the polynomials $\Phi_n$ on the arc of unit disc defined by $\alpha_{\Lambda_b\Lambda^*}$, we find
\begin{equation}
\begin{aligned}
    \rho_0 =&\, 0.572048\,, & \rho_1 =& -0.624505\,, & \rho_2 =& 0.641153 \,,\\
    \rho_3 =& -0.647652\,, & \rho_4 =&\, 0.650567 \,, & \rho_5 =&\, -0.652072\,.
\end{aligned}
\end{equation}
The orthonormal polynomials $p_n$ then take the form
\begin{equation}
    p_n = \frac{\Phi_n}{N_n}\,, \quad \mathrm{and} \qquad N_n = \left[2\alphaLbLst \prod_{i=0}^{n-1} (1-\rho_i^2)\right]^{1/2}\,.
\end{equation}

\bibliographystyle{utphys}
\bibliography{references}

\end{document}